\pdfoutput=1
\PassOptionsToPackage{unicode}{hyperref}
\documentclass[11pt,final]{article}

\usepackage{authblk}
\usepackage[utf8]{inputenc}
\usepackage[T1]{fontenc}
\usepackage[hyphens]{url}

\usepackage[english]{babel}
\usepackage[activate=true,final,kerning=true,babel=true,protrusion=true,tracking=true]{microtype}
\usepackage{zi4}

\usepackage[backend=bibtex,style=numeric]{biblatex}
\usepackage{mathtools,amsfonts,amssymb,mathabx}
\usepackage{csquotes}

\usepackage{tikz}
\usepackage{varioref}
\usepackage[pdfpagelabels=true]{hyperref}
\usepackage{cleveref}
\usepackage{xcolor}
\usepackage{adjustbox}
\usepackage{caption}
\usepackage{subcaption} 
\usepackage{listings}
\usepackage[symbol]{footmisc}
\usepackage{enumitem}
\usepackage{lmodern}
\usepackage{cprotect}

\newcommand*{\ctr}[1]{~\hfill #1 \hfill~}

\hypersetup{
	linktoc = page,
	pdfpagemode = UseNone,
	colorlinks,
	linkcolor={red!50!black},
	citecolor={blue!50!black},
	urlcolor={blue!80!black}
}

\begin{document}
\date{}

\author[1]{Hadrien Barral} 
\author[1,5]{Georges-Axel Jaloyan}
\author[2,4]{Fabien Thomas-Brans}
\author[3]{Matthieu Regnery}
\author[1,6]{Rémi Géraud-Stewart\footnote{This work was performed when R\'emi G\'eraud was at \'Ecole normale supérieure.}}
\author[2,1]{Thibaut Heckmann}
\author[3,1]{Thomas Souvignet}
\author[1]{David Naccache}

\affil[1]{DIENS, \'Ecole normale supérieure, CNRS, PSL Research University, Paris, France}
\affil[2]{Centre de recherche de l'école des officiers de la Gendarmerie nationale, Melun, France}
\affil[3]{\'Ecole des sciences criminelles, Université de Lausanne, Suisse}
\affil[4]{Université de Limoges, Xlim, Limoges, France}
\affil[5]{Secure-IC S.A.S., Paris, France}
\affil[6]{QPSI, Qualcomm Inc., San Diego CA, USA}

\title{A forensic analysis of the Google Home: repairing compressed data without error correction}

\maketitle

\thanks{Modified version of our paper that appeared at Forensic Science International: Digital Investigation.}

\begin{abstract}
This paper provides a detailed explanation of the steps taken to extract and repair a Google Home's internal data. Starting with reverse engineering the hardware of a commercial off-the-shelf Google Home, internal data is then extracted by desoldering and dumping the flash memory. As error correction is performed by the CPU using an undisclosed method, a new alternative method is shown to repair a corrupted SquashFS filesystem, under the assumption of a single or double bitflip per gzip-compressed fragment. Finally, a new method to handle multiple possible repairs using three-valued logic is presented.
\end{abstract}

\pagebreak
\section{Introduction}

The widespread adoption of embedded devices, especially in the context of the \emph{Internet of Things} has led to user data being scattered across multiple platforms with limited computational capabilities.
Forensic analysis frequently assumes the ability to extract data from those embedded systems, that may be used as evidence in courts.

Among such devices are smart speakers, allowing users to control a whole range of services with their voices -- \emph{e.g.}, music, news, calendar -- as well as smart home appliances that may be connected to the speaker -- \emph{e.g.}, lights and temperature control, kitchen appliances, home access security. Forensic analysis of smart speakers can be of critical importance, due to their strategic position in any house.

First introduced in 2016, the Google Home is a popular smart speaker developed by Google. Its original version features a cylindrical case with voice and touch-pad inputs, connecting to various devices through Wi-Fi or Bluetooth. Several variants have since been released, including the Google Home mini or the discontinued Google Home max.

Although a significant part of its software is open sourced and its data stored on the cloud, the process of recovering a smart speaker's internal data is still needed in some cases, mostly to check data authenticity, and collect unique identifiers aimed at efficiently requesting missing data from providers. As such, hardware based techniques are the most reliable and reproducible.

However, such methods may be hampered by data scrambling (among which encoding or encryption), hardware damage, or errors introduced during the reading process. Depending on the amount and type of error, large chunks of data may be corrupted and thus unusable to further investigation. This is emphasized by the ubiquitous usage of compression and or encryption in embedded devices.

The contributions of this paper can be summarized as follows:
\begin{enumerate}[nolistsep,before=\vspace{0.5\baselineskip},after=\vspace{0.5\baselineskip}]
	\item This paper details the various steps taken to reverse engineer part of the Google Home's hardware and dump its flash memory.
	\item This paper explains the analysis of the data obtained from the dump, and the identification of corrupted compressed sections of the flash.
	\item This paper introduces a new method to repair corrupted SquashFS filesystems, under the assumption of a single or double bitflip model per fragment.
	\item This paper shows this technique applied to the Google Home, and a method using three-valued logic to merge multiple possible repairs.
\end{enumerate}

The paper is organized as follows. Section~\ref{sec:related} surveys the related work performed on similar devices. Section~\ref{sec:ghome} introduces the target platform under study, and describes the steps taken at identifying and dumping its memory. Section~\ref{sec:dumping} analyzes and identifies the corrupted SquashFS filesystem, that is repaired in Section~\ref{sec:repair}. Section~\ref{sec:results} describes and discusses the obtained results, followed by a short conclusion in Section~\ref{sec:conclusion}.

\section{Related work}
\label{sec:related}

Smart speakers have been at the center of several forensics research efforts. Chung \emph{et al.}~\cite{CHUNG2017} studied the Amazon Alexa, by looking at the various API calls and local data stored by the companion smartphone application linked to the smart speaker. Engelhardt~\cite{ENGELHARDT2019} extended the methodology to the Google Home's companion application on Samsung Galaxy S5 Android phone.

Hardware based approaches in the context of forensics analysis have been performed on the Amazon Echo by Youn \emph{et al.}~\cite{YOUN2021} in 2021. In 2019, Qian \emph{et al.} presented an attack on the Google Home through a vulnerability in SQLite and curl~\cite{QIAN2019}. As a first step in their reverse engineering efforts, they dumped the device's memory to recover the firmware, though no details were given on the correction of bitflips during the dump. In 2020, Courk~\cite{COURK12020,COURK22020} presented another attack on the Google Home, with similar initial reverse engineering steps taken. Courk managed to guess the SoC's error correction algorithm, by trial and error method. Although impressive, their results are unfortunately only limited to the particular SoC that equips this specific device. Google Home features several SoC that are not guaranteed to use the same \emph{error correcting code} (ECC) algorithm. In what follows, another repair method for gzip compressed data is presented, that does not rely on finding the ECC algorithm, assuming a low error rate.

Repairing compressed data has not been subject to significant research efforts, with tools mostly focused on extracting uncorrupted fragments from corrupted archives. A method to recover parts of a corrupted file compressed with DEFLATE has been published by Park \emph{et al.} in 2008~\cite{PARK2008}. Their method leverages the Huffman coding in DEFLATE to drop prefixes of unrepairable data until the corrupted area does not influence the rest of the file, yielding several chunks of uncorrupted data. Another work published by Wang \emph{et al.} in 2019~\cite{WANG2019} explains how to modify the LZSS compression algorithm to add redundancy with minimal performance impact, without modifying the decompressing algorithm. Note that error reparation must still be performed with a custom algorithm. Unlike existing work, the method presented in this paper aims at repairing the corrupted parts of third-party data, relying solely on already existing redundancy embedded therein.

\section{Google Home hardware}
\label{sec:ghome}

\begin{figure}[!b]
	\centering
	\includegraphics[width=\textwidth]{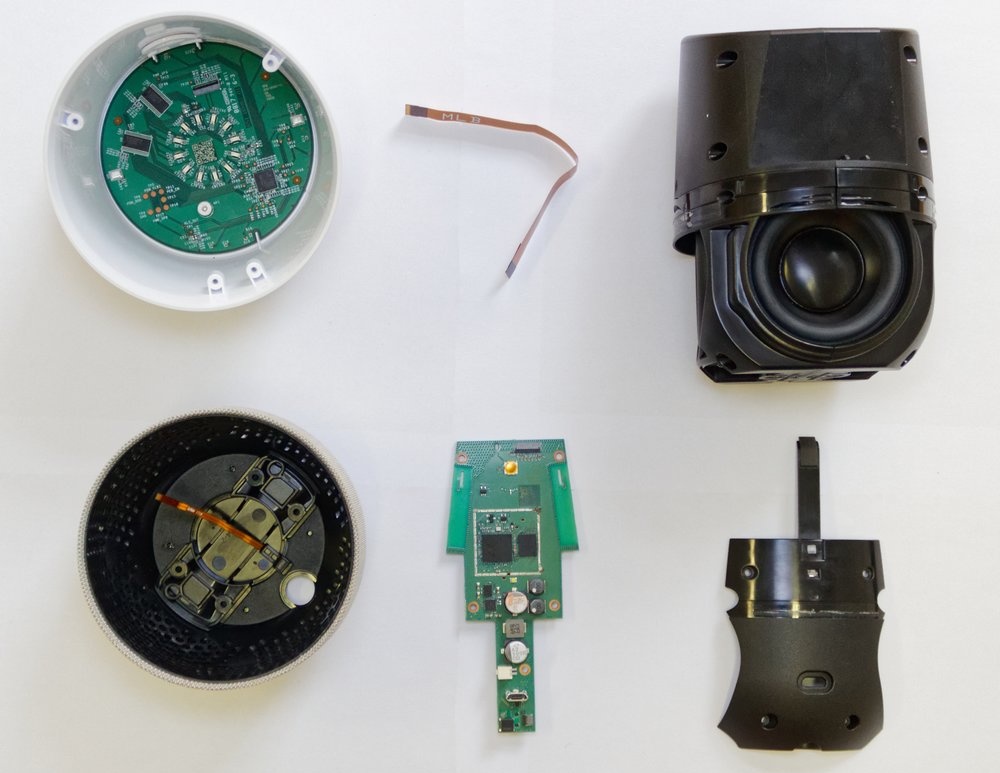}
	\captionsetup{singlelinecheck=off}
	\caption[LoF entry]{Google Home main components (from top-left in clockwise order):
		\begin{itemize}[noitemsep,nolistsep]
			\item IO board, glued to the case.
			\item 16-pin ribbon cable connecting the two boards.
			\item Speaker assembly.
			\item Cover for the bottom board.
			\item Bottom board.
			\item Bottom of the case.
	\end{itemize}}
	\label{fig:global}
\end{figure}

This section discusses the steps taken to identify all components on the system, aiming at determining which components contain stored data. The goal is to extract the firmware to analyze its contents.

In the forensic or security fields, firmware analysis is an important part of understanding internal mechanisms. From the firmware, it is possible to understand how the equipment works and how user data is collected, stored, encrypted or sent.

The equipment chosen as a target is a Google Home in its original variant, built in June 2017. Some tutorials, on open access on the Internet, are available to help with the teardown~\cite{HARVARD2016}. It is quite easy to extract the electronic boards constituting the system, in Fig.~\ref{fig:global}.

\begin{figure}[!t]
	\centering
	\includegraphics[width=\textwidth]{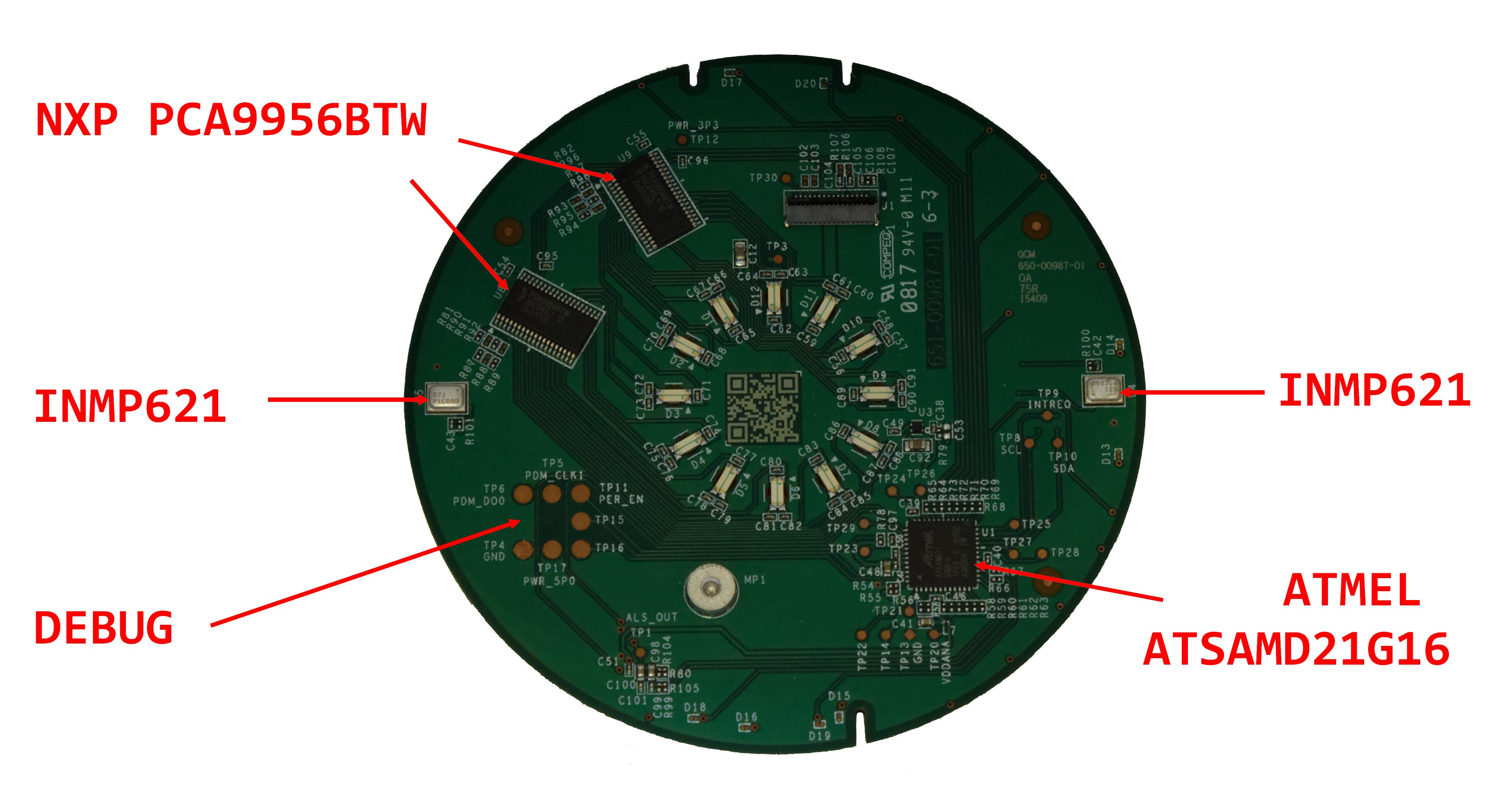}
	\captionsetup{singlelinecheck=off}
	\caption[LoF entry]{View of the backside of the IO board (main components and test points are highlighted). Particularly, the following components have been identified:
		\begin{itemize}[noitemsep,nolistsep,before=\vspace{0.15\baselineskip},after=\vspace{0.15\baselineskip}]
			\item Two NXP PCA9956BTW led controllers~\cite{NXP2020}.
			\item Two INMP621 microphones~\cite{INMP2014}.
			\item One ATMEL ATSAMD21-G16 Cortex M0+ microcontroller~\cite{MICROCHIP}.
	\end{itemize}}
	\label{fig:capacitive}
\end{figure}

\begin{figure}[!t]
	\centering
	\begin{subfigure}{0.49\textwidth}
		\centering
		\includegraphics[width=\linewidth]{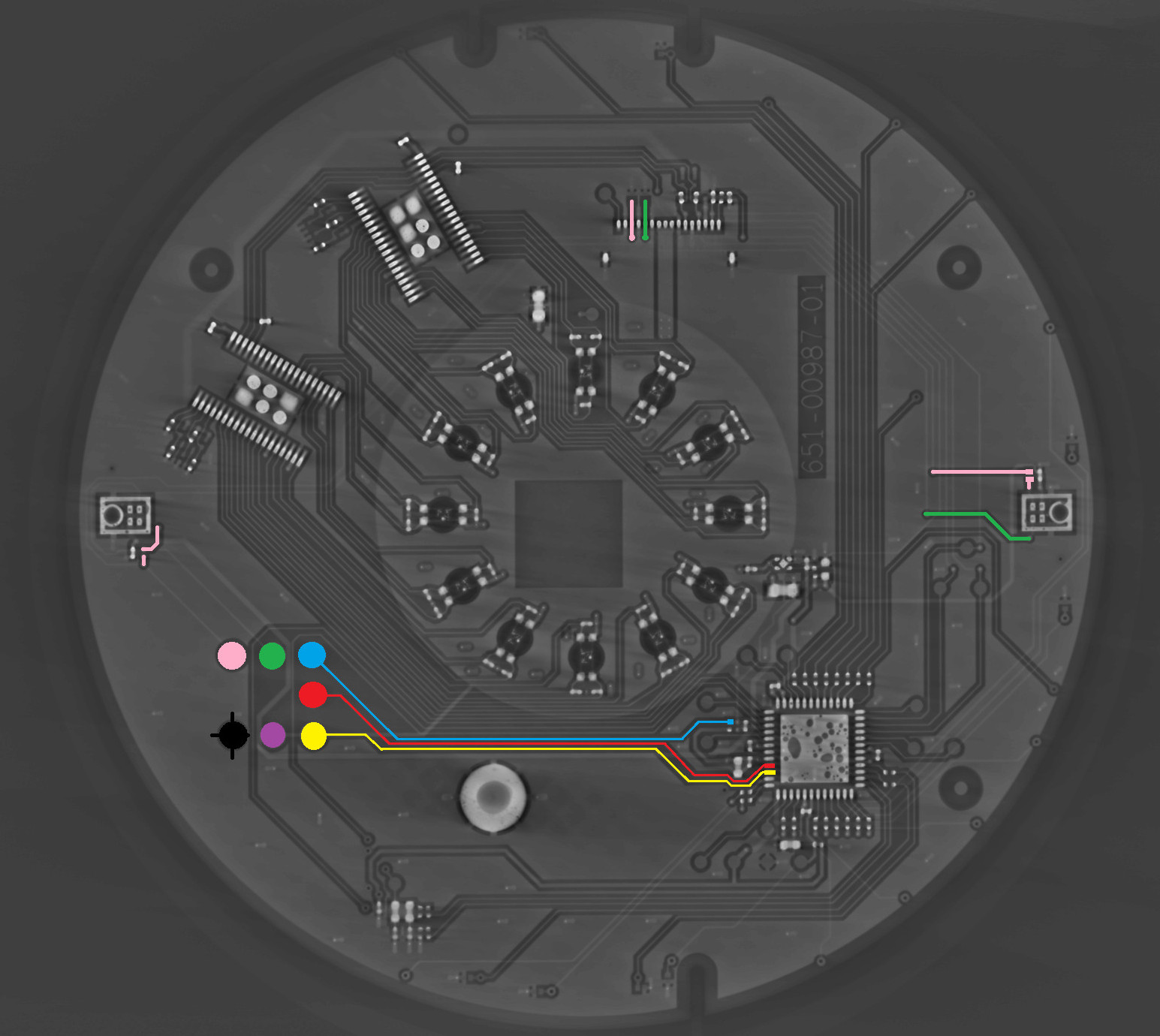}
		\captionsetup{singlelinecheck=off}
		\caption{1st layer. }
	\end{subfigure}
	\begin{subfigure}{0.49\textwidth}
		\centering
		\includegraphics[width=\linewidth]{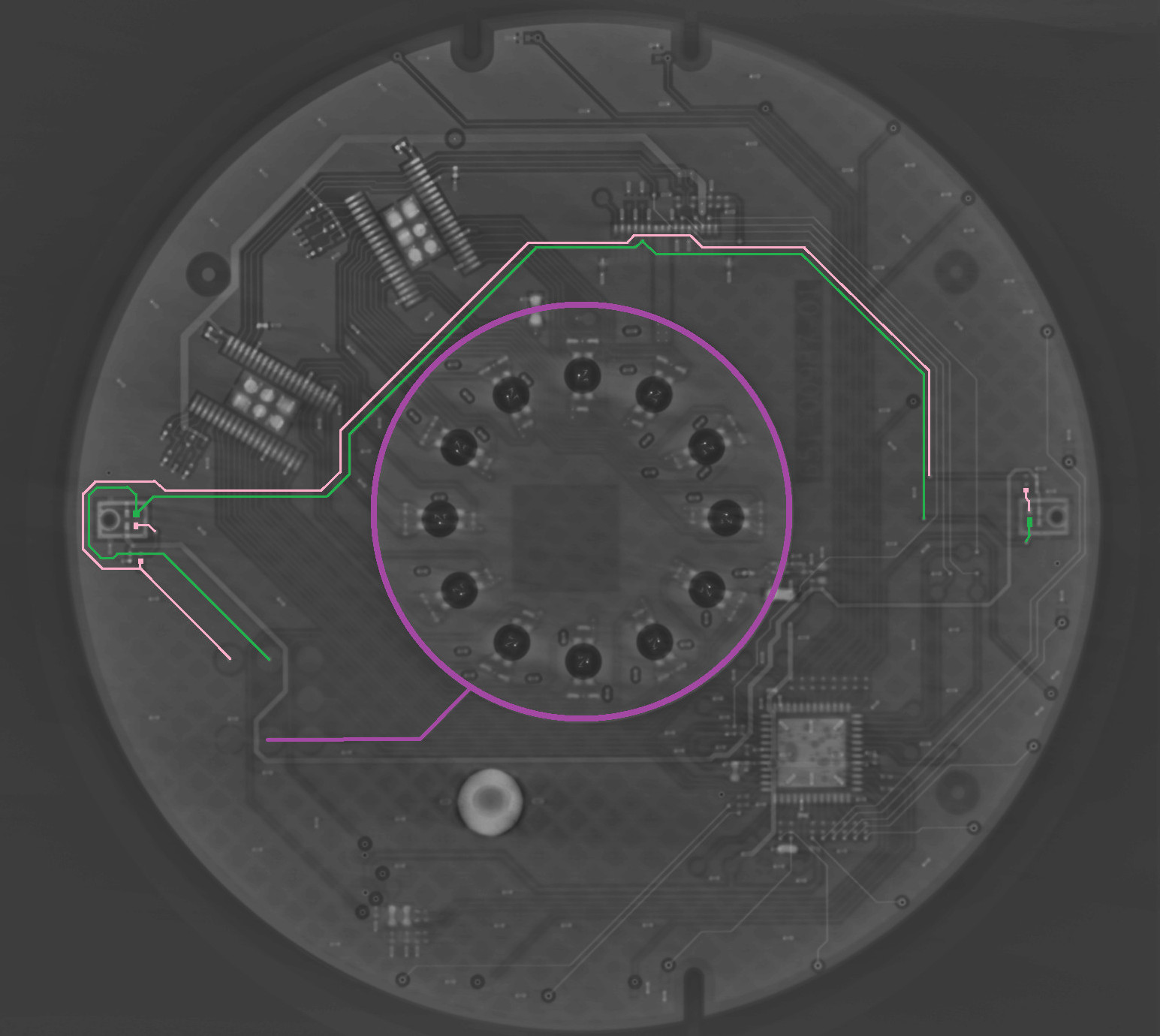}
		\captionsetup{singlelinecheck=off}
		\caption{2nd layer. }
	\end{subfigure} \\
	\begin{subfigure}{0.49\textwidth}
		\centering
		\includegraphics[width=\linewidth]{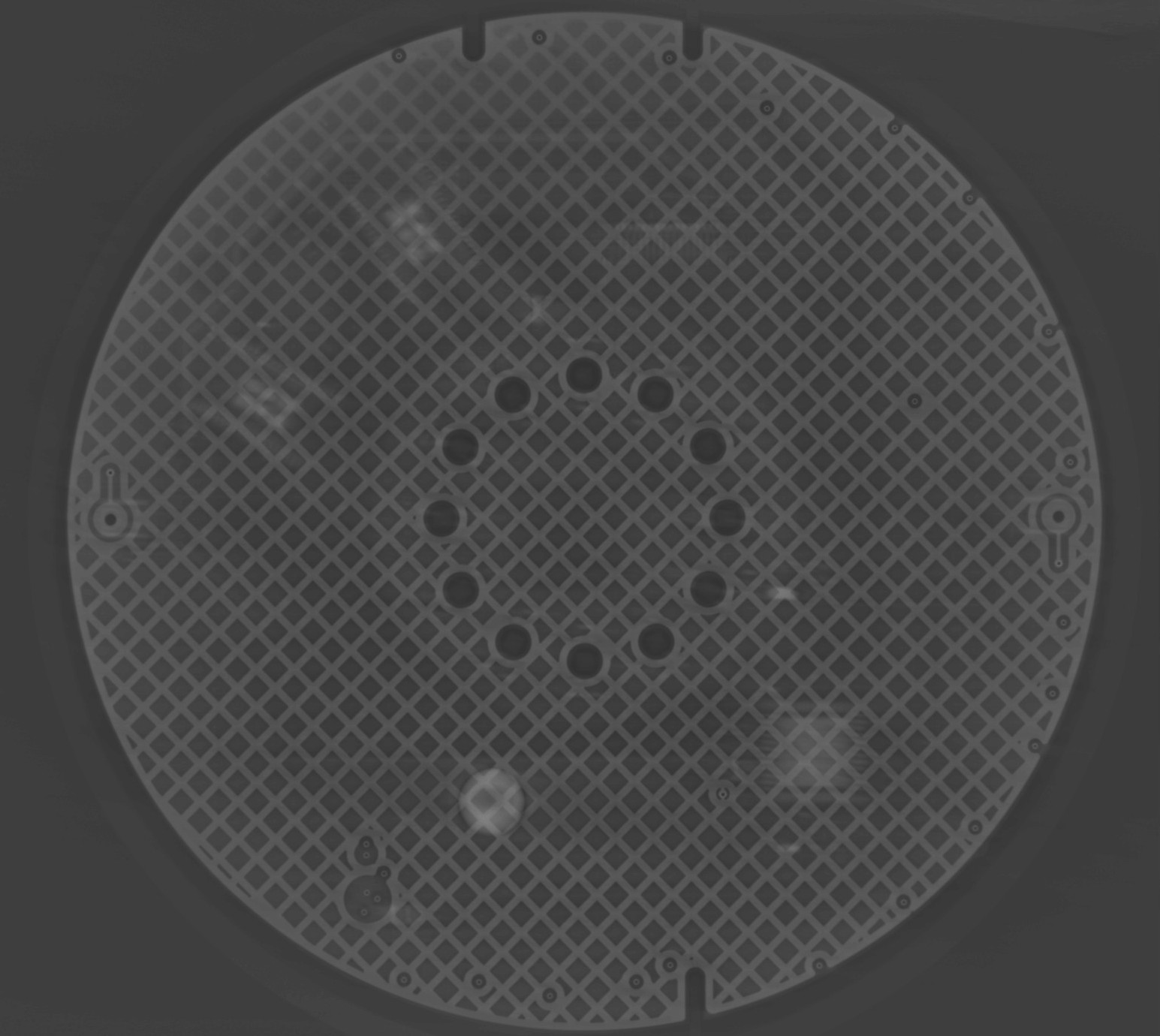}
		\captionsetup{singlelinecheck=off}
		\caption{3rd layer. \\ \strut }
	\end{subfigure}
	\begin{subfigure}{0.49\textwidth}
		\centering
		\includegraphics[width=\linewidth]{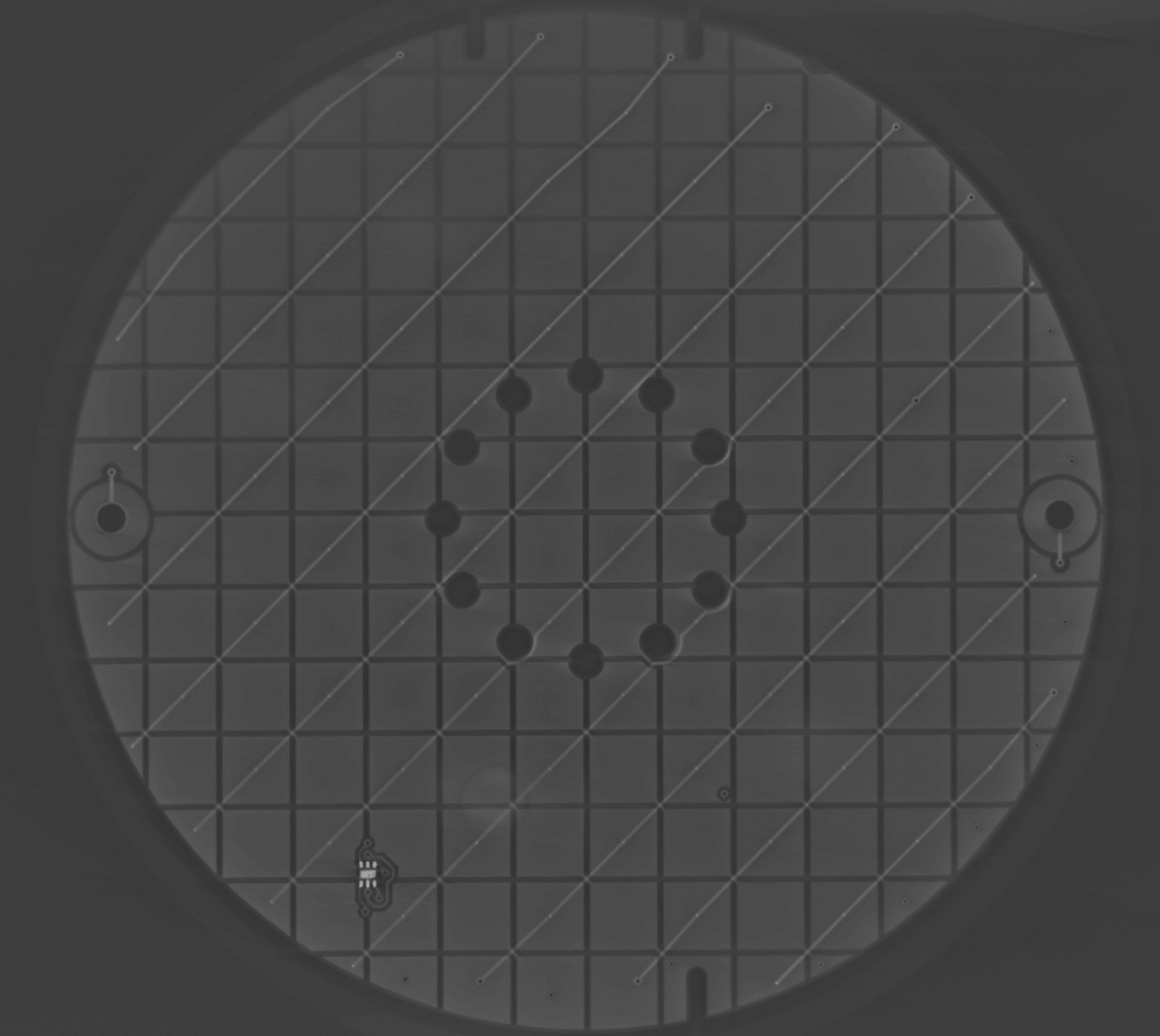}
		\captionsetup{singlelinecheck=off}
		\caption{4th layer. This side is glued to the casing.\strut}
	\end{subfigure}
	\caption{X-Ray view the top IO board. The PCB is made of four layers, each yielding a 2D picture. The seven test points visible in Fig.~\ref{fig:capacitive} are traced up to the respective components.}
	\label{fig:IOXray}
\end{figure}

The first board -- called the IO board --, located at the top part of the Google Home has a circular shape (Fig.~\ref{fig:capacitive}). The \emph{Printed Circuit Board} (PCB) has 4 layers, shown in Fig.~\ref{fig:IOXray} using X-Ray tomography. The topside (layer 4) is composed of 12 LEDs and a capacitive grid array acting as a touchscreen. This side is glued to the upper case and thus cannot be accessed without destroying the product. The X-Ray tomography shows that this side does not have components which can contain data.

The backside of the IO board has \emph{active components} -- components controlling signals in the circuit which are often based on transistors. Two of those active components are led controllers and the two others are microphones. Last active component is an ATMEL Cortex M0+ microcontroller containing 64 KiB internal flash memory. This memory has a limited capacity and is generally dedicated for small bare-metal code to be executed rather than full-fledged firmware.

On the board, it is also possible to observe seven test points (with pins named respectively from top-leftmost in clockwise order: \texttt{TP6 PDM\_DO0}, \texttt{TP5 PDM\_CLK1}, \texttt{TP11 PER\_EN}, \texttt{TP15}, \texttt{TP16}, \texttt{TP17 PWR\_SP0}, \texttt{TP4 GND}). Other test pins are available, that can be linked to various components using the datasheet, not detailed in this paper. Test points are placed by designers in order to help in the conception of the board: they can have several diagnostic functions like checking the values of certain signals or programming a component after soldering it on the board.

For programming and debugging operations, several protocols are used, but the most commons are the JTAG~\cite{IEEEJTAG} or SWD~\cite{ARMSWD} protocols. By looking at the ATMEL microcontroller's datasheet, some pads correspond to an SWD bus. The SWD protocol requires two power (\texttt{GND} and \texttt{PWR}) and two data wires (\texttt{SWDIO} and \texttt{SWCLK}). The \texttt{GND} and \texttt{PWR} wires allow respectively to share a common ground and power line between the programmer and the board. Some test points on the board may thus correspond to an SWD interface. The first is \texttt{SWDIO} ensuring data exchanges between the programmer and the microcontroller. The second wire is \texttt{SWCLK}, which is a clock.

It is possible to identify each test pin by partially reversing the PCB with a multimeter in continuity test mode. This process can be refined by tracing the pins using X-Ray tomography (Fig.~\ref{fig:IOXray}) to confirm the role of TP15 as \texttt{SWCLK}, TP16 as \texttt{SWDIO}, TP4 as \texttt{GND}, TP17 as \texttt{PWR} and TP11 as (pulled-up) \texttt{RESET}. The pins TP6 and TP5 expose the sound recorded by the microphones, in the form of a digital signal using \emph{Pulse-density modulation}, with TP6 being the data, and TP5 the clock signal.

\begin{figure}[tp]
	\centering
	\begin{subfigure}{\textwidth}
		\centering
		\includegraphics[width=\textwidth]{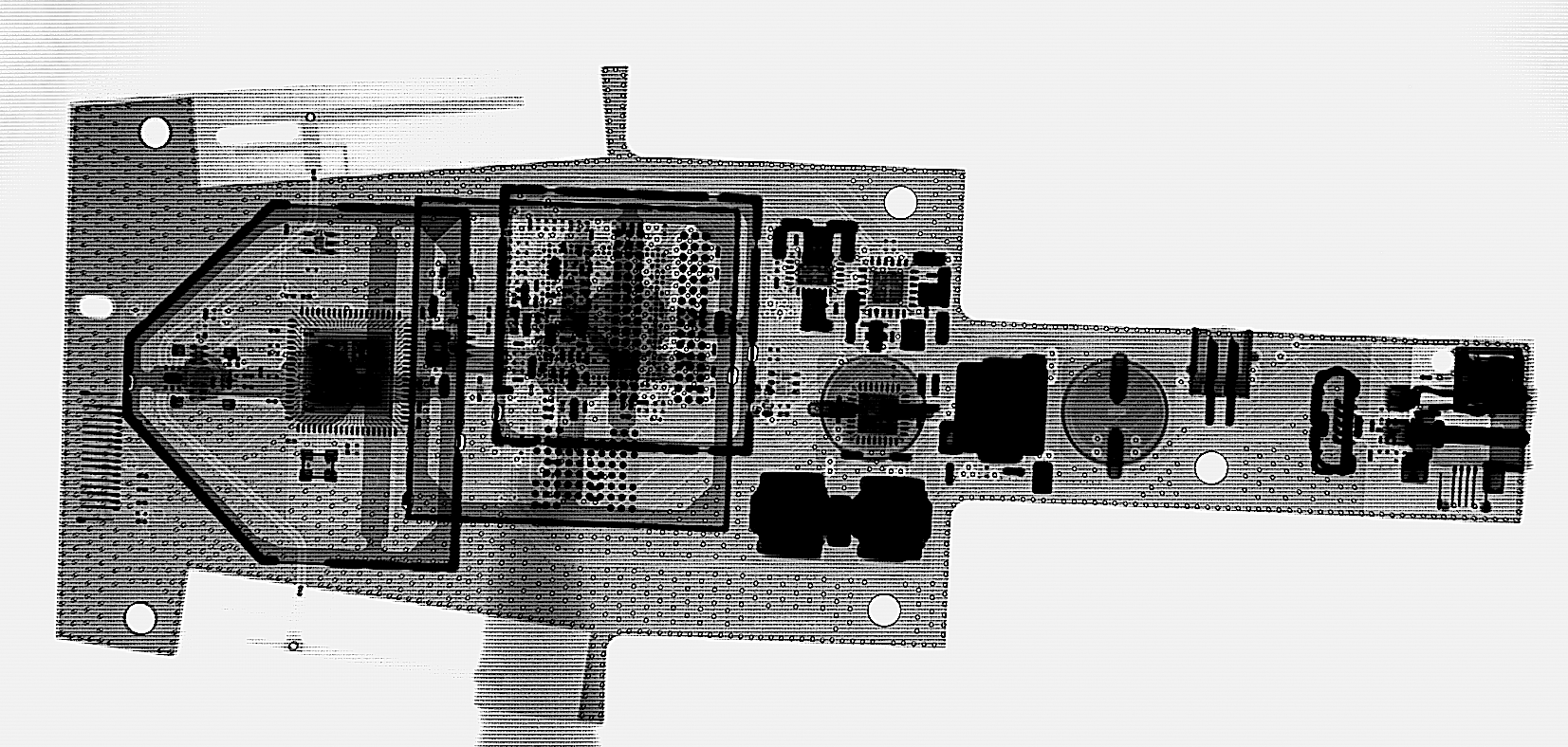}
		\caption{2D X-Ray view of the bottom board. Active components can be identified behind the RF shields.}
		\vspace{0.5cm}
		\label{fig:tomography}
	\end{subfigure}
	\begin{subfigure}{\textwidth}
		\centering
		\includegraphics[width=\textwidth]{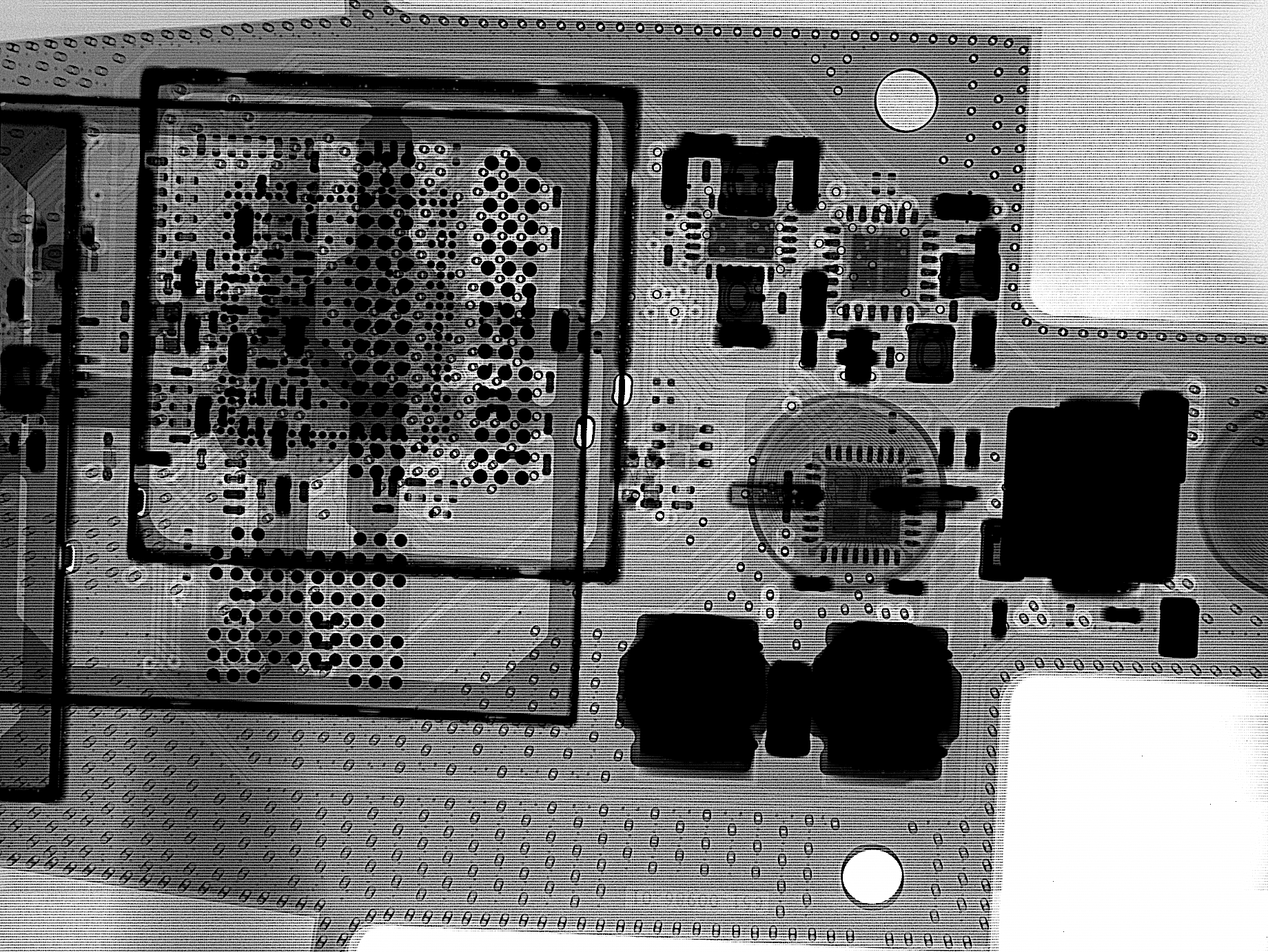}
		\caption{2D X-Ray close-up view of the bottom board. The RAM chip, the flash memory and the CPU are located behind the RF shields.}
		\label{fig:tomography2}
	\end{subfigure}%
	\caption{X-Ray view of the bottom board.}
\end{figure}

The top board is linked to the bottom board with a 16-pin ribbon cable. The second board features several RF shields hiding the components beneath. A 2D X-Ray view is performed to locate the components under the shields (Fig.~\ref{fig:tomography}). With this view, the nature of the components and whether they are active or passive can be determined without depackaging the chips. As a reminder, data is contained in active components. The 2D X-Ray view also allows to identify components that may be located at the shields' edge. If components are located too close to the shield, there is a risk of damaging them when removing the shield.

\begin{figure}[!tp]
	\centering
	\begin{subfigure}{0.95\textwidth}
		\includegraphics[width=\linewidth]{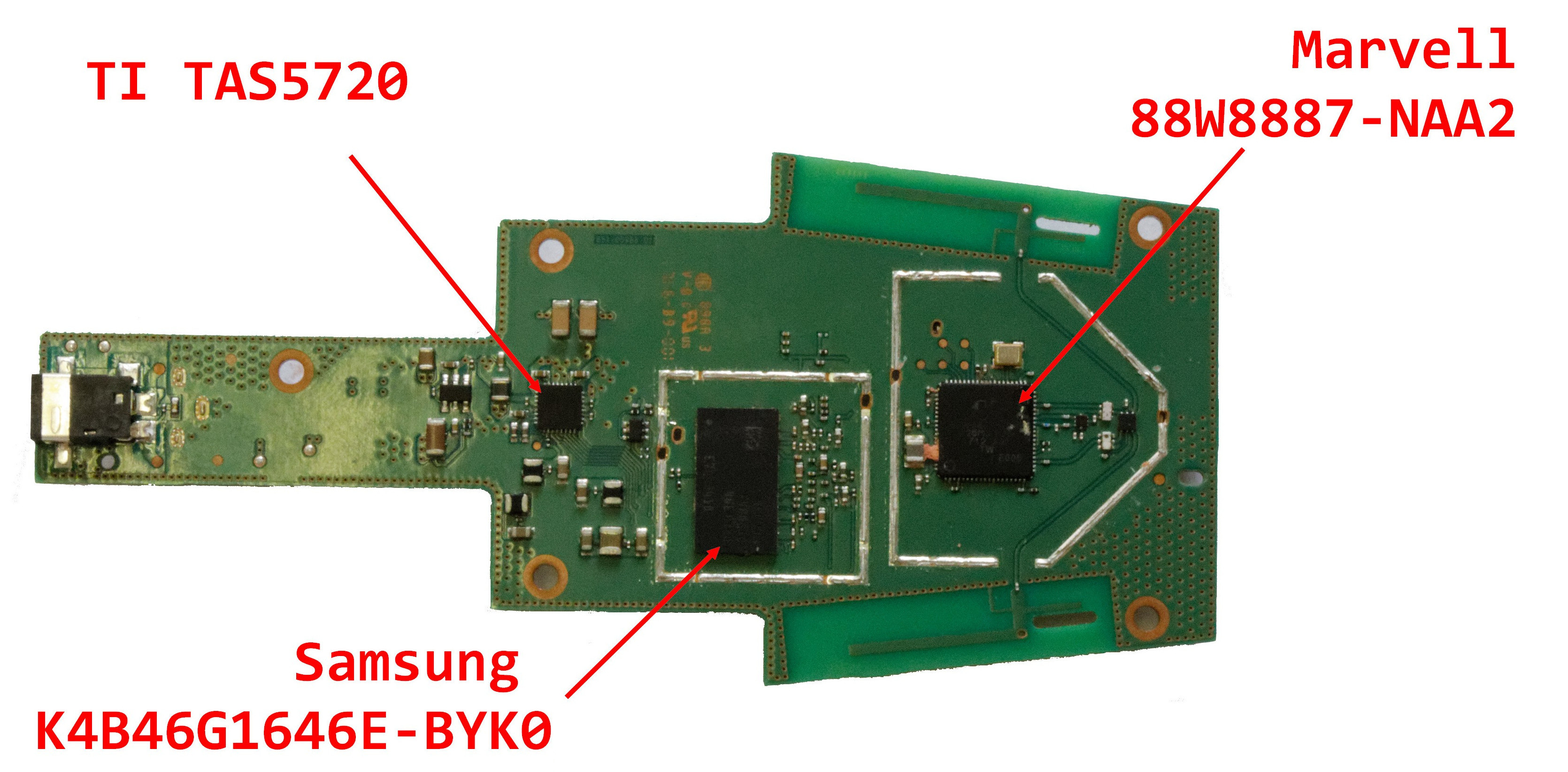}
		\captionsetup{singlelinecheck=off}
		\caption{Topside view of the main board (main components are highlighted).
			\begin{itemize}[noitemsep,nolistsep,after=\vspace{\baselineskip}]
				\item Marvell Avastar 88W8887-NAA2 WiFi, Bluetooth and NFC communication chip~\cite{MARVELLAVASTAR}.
				\item Texas Instruments TAS5720 audio amplification chip~\cite{TI2015}.
				\item Samsung K4B4G1646E-BYK0 512MB DDR3 SDRAM chip~\cite{SAMSUNGDDR}.
			\end{itemize}
		}
		\label{fig:maindown}
	\end{subfigure} \\
	\begin{subfigure}{0.95\textwidth}
		\includegraphics[width=\linewidth]{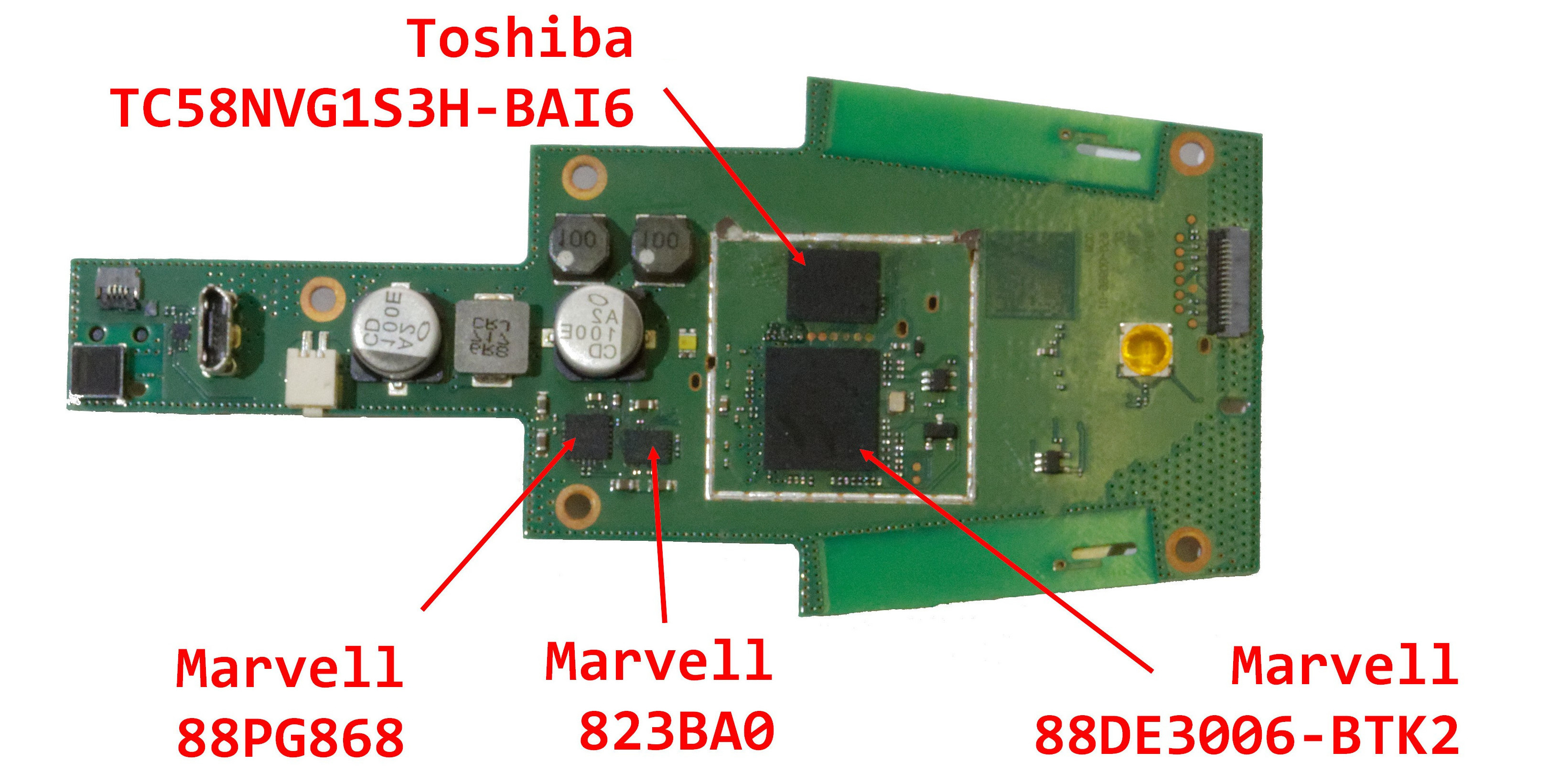}
		\captionsetup{singlelinecheck=off}
		\caption{Backside view of the main board (main components are highlighted).
			\begin{itemize}[noitemsep,nolistsep,after=\vspace{\baselineskip}]
				\item Marvell 88PG868 I2C DC/DC voltage regulator~\cite{MARVELL2016}.
				\item Unidentified Marvell 823BA0 chip.
				\item 88DE3006-BTK2: 2-core ARM Cortex-A7 Marvell Armada 1500 Mini Plus~\cite{GOOGLEKERNEL}.
				\item TC58NVG1S3H-BAI6 Toshiba NAND256MB flash memory~\cite{TOSHIBANAND}.
		\end{itemize} }
		\label{fig:mainup}
	\end{subfigure}%
	\caption{Optical view of the main board with active components highlighted, RF shields have been removed.}
\end{figure}

There are several solutions for removing the shields while minimizing the risk of damaging the surrounding components. One solution is to desolder the shield with a machine (\emph{i.e.}, ZEVAC or PDR). Another solution is to process the upper part of the shield with a micro-milling machine (\emph{i.e.}, precision sander or Dremel). When using a milling machine, it is necessary to use a small diameter milling cutter, in order to be precise and to avoid touching other elements. The most efficient way is to machine the edge of the shields, without going too deep to avoid overshooting. The last operation consists in sliding a strong scalpel blade to break the thin layer of metal remaining. Under no circumstances should a shield be ripped off with pliers, as these are connected to the ground, and there will be a risk of tearing off the ground plane.

\begin{figure}
	\centering
	\begin{subfigure}{0.49\textwidth}
	\centering
    \includegraphics[width=\textwidth]{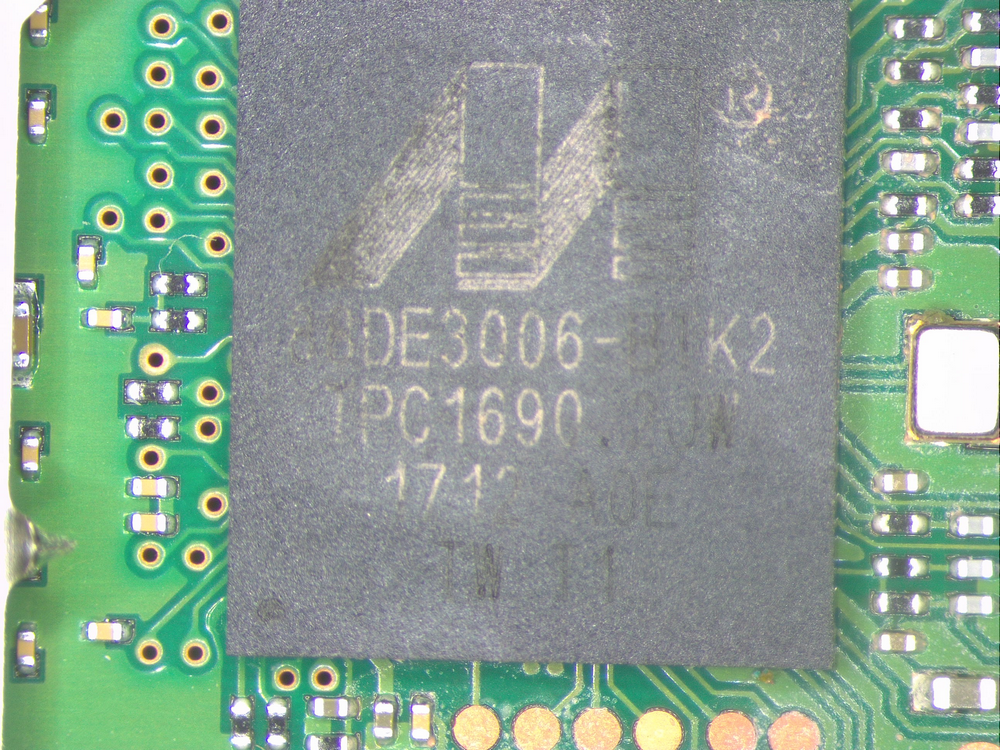}
	\end{subfigure}
\begin{subfigure}{0.49\textwidth}
    \centering
    \includegraphics[width=\textwidth]{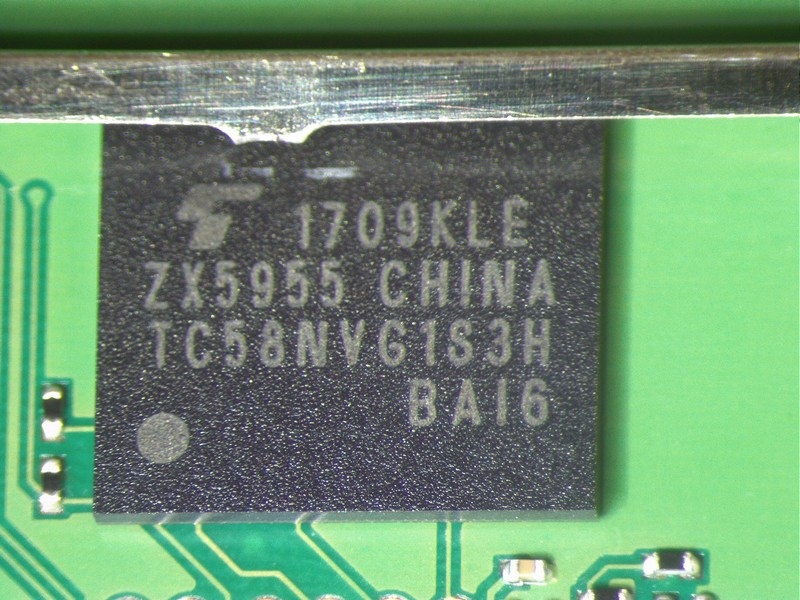}
    \end{subfigure}
    \caption{Close up on the Marvell Armada 1500 Mini Plus SoC (reference: 88DE3006-BTK2) and its Toshiba 256MB NAND flash memory (reference: TC58NVG1S3H-BAI6). Six test points can be seen on the bottom of the left image.}
    \label{fig:closeup}
\end{figure}

After removing the \emph{electromagnetic shields} (or \emph{RF shields}), it is possible to identify the active components. On the backside (Fig.~\ref{fig:maindown}), there is a WiFi, Bluetooth and NFC communication chip. Next to it, there is a component for audio amplification for the speakers and a Samsung 512MB DDR3 SDRAM chip. On the topside (Fig.~\ref{fig:mainup}), there is an I2C DC/DC voltage regulator and an unidentified chip. The two interesting chips are the \emph{System on Chip} (SoC), which is a 2-core ARM Cortex-A7 Marvell Armada 1500 Mini Plus~\cite{GOOGLEKERNEL} connected to a Toshiba NAND256MB flash memory~\cite{TOSHIBANAND} (shown in Fig.~\ref{fig:closeup}). The flash memory being the only component that can contain a significant amount of data (\emph{e.g.}, an android image), the study will focus on it.

\begin{figure}[tp]
	\centering
	\begin{subfigure}{0.45\textwidth}
		\centering
		\includegraphics[width=\linewidth]{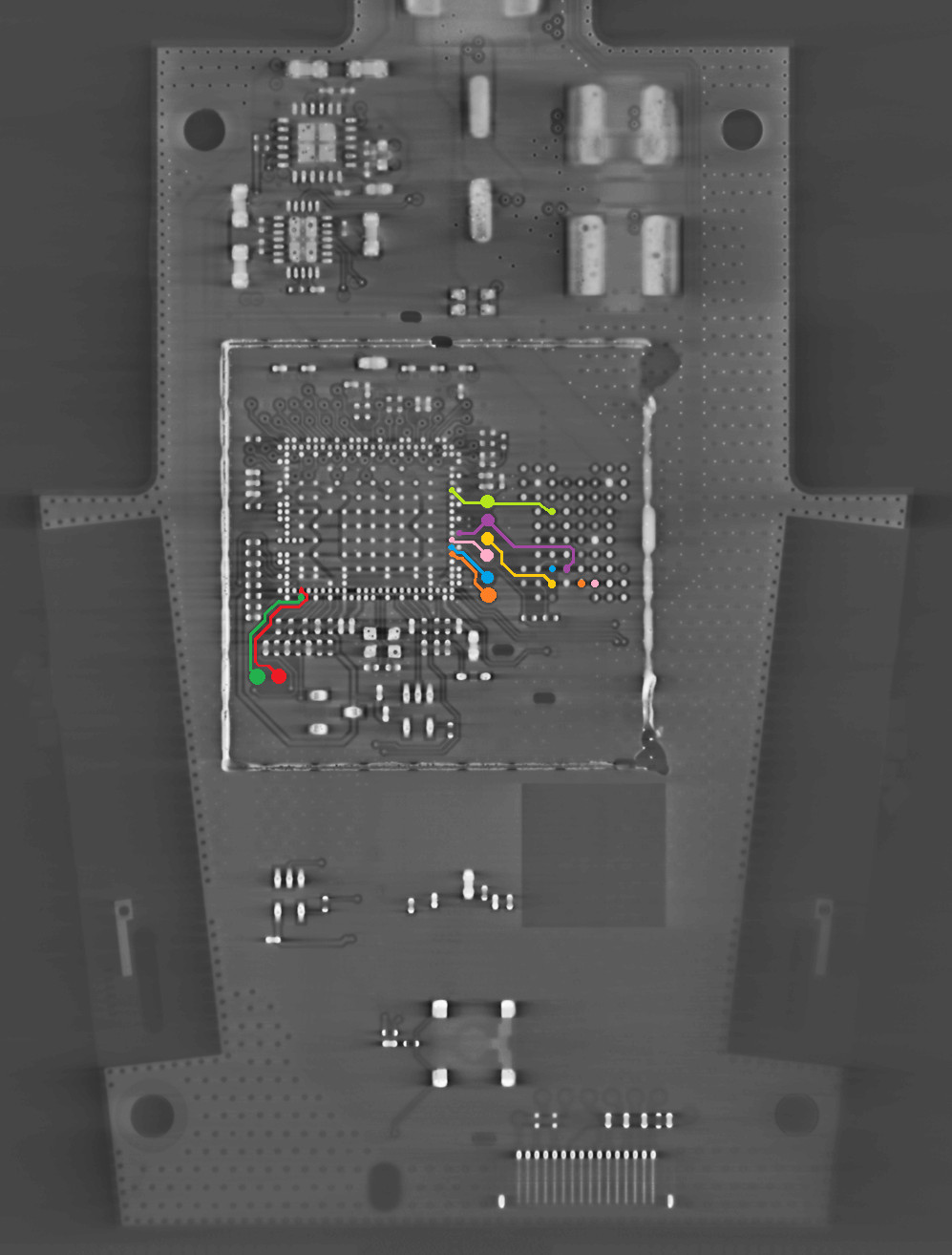}
		\captionsetup{singlelinecheck=off}
		\caption{1st layer. }
	\end{subfigure}
	\begin{subfigure}{0.45\textwidth}
		\centering
		\includegraphics[width=\linewidth]{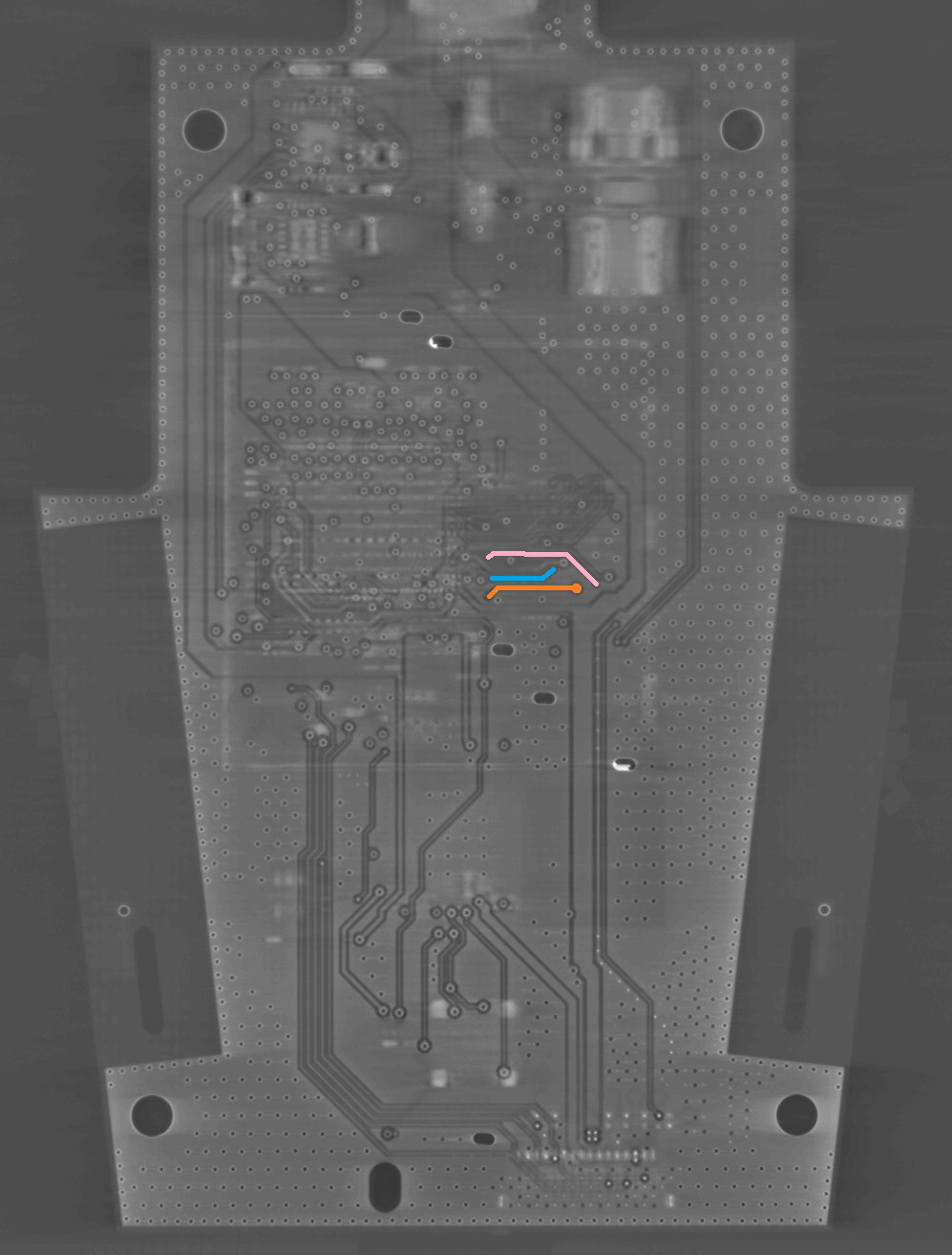}
		\captionsetup{singlelinecheck=off}
		\caption{2nd layer. }
	\end{subfigure} \\
	\begin{subfigure}{0.45\textwidth}
		\centering
		\includegraphics[width=\linewidth]{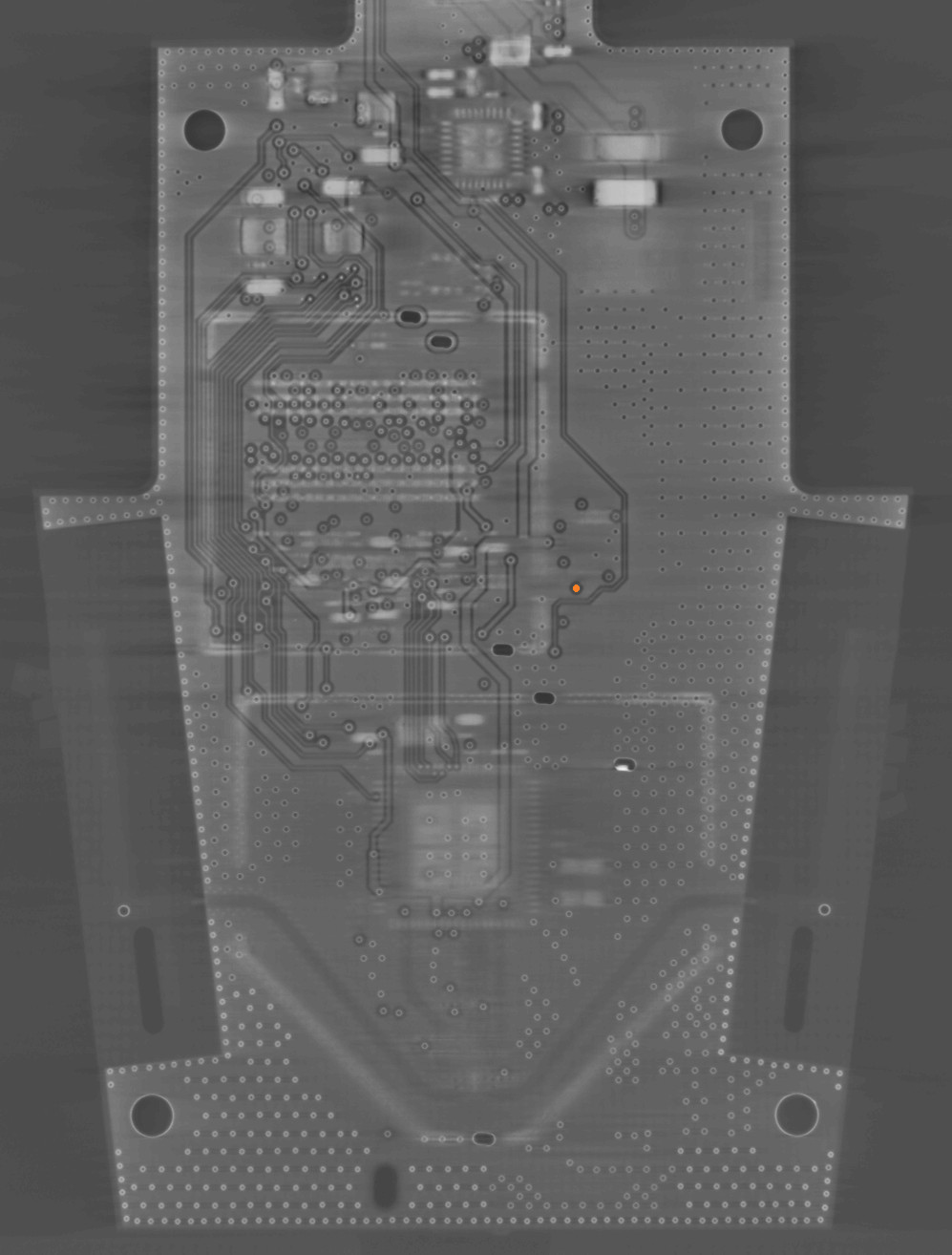}
		\captionsetup{singlelinecheck=off}
		\caption{3rd layer. }
	\end{subfigure}
	\begin{subfigure}{0.45\textwidth}
		\centering
		\includegraphics[width=\linewidth]{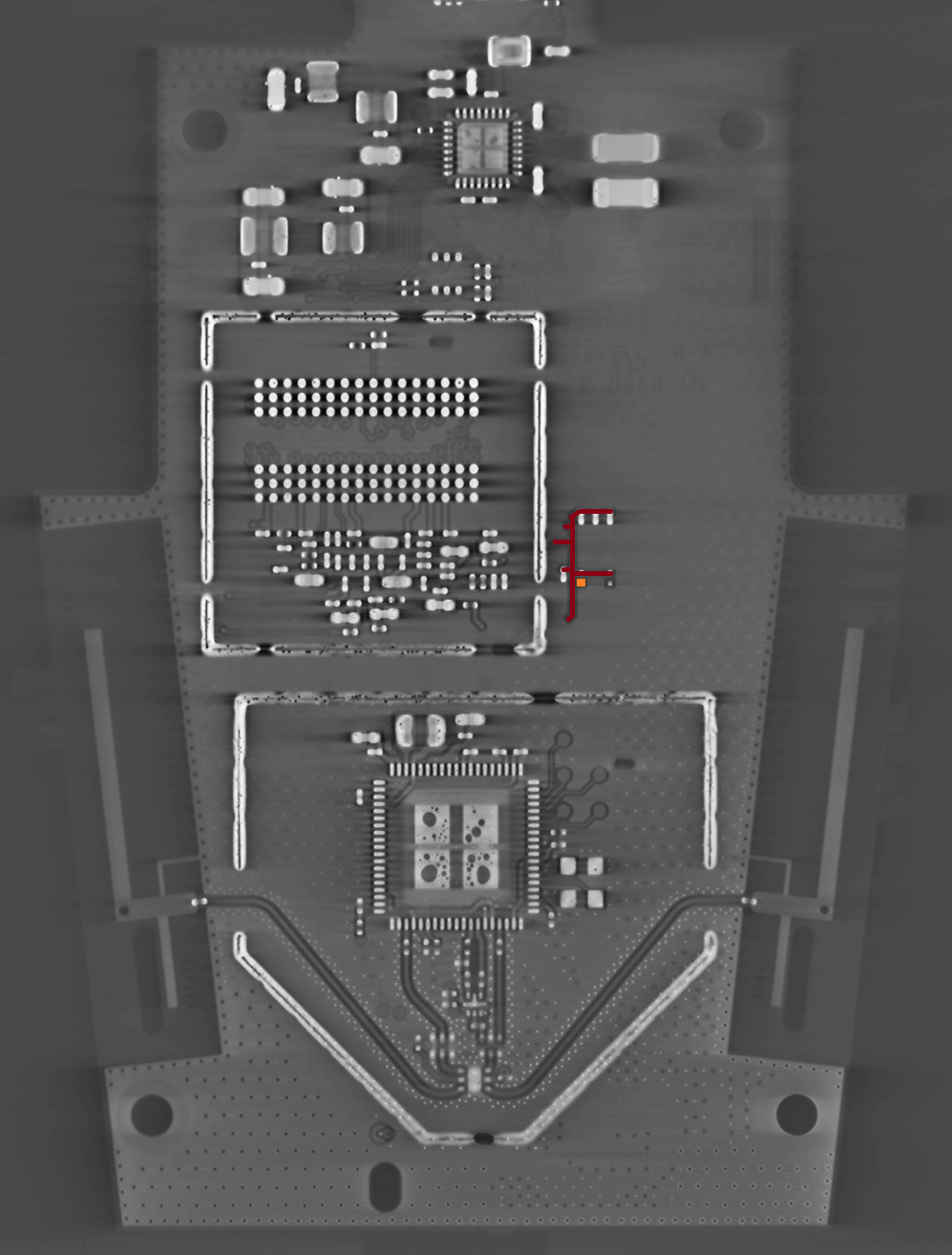}
		\captionsetup{singlelinecheck=off}
		\caption{4th layer. Red trace is VCC.}
	\end{subfigure}
	\caption{X-Ray view of the bottom main board. The PCB is made of four layers, each yielding a 2D picture. The six test points are traced up to the respective components. Two other tests points have been traced to unidentified pins of the SoC. The red trace in layer 4 is connected to VCC and is used to pull-up the orange test point (\texttt{CE}).}
	\label{fig:MainXray}
\end{figure}

According to the datasheet, the Toshiba memory is using an Single Data Rate (SDR) protocol respecting the ONFI standard~\cite{ONFI}. This means that the number of signals to drive is seventeen: seven for controls, eight for Input/Output, and two for \texttt{VCC} and \texttt{GND}.

Several test pins are available on the main board. An X-Ray tomography in Fig.~\ref{fig:MainXray} allows to trace six test pins located between the SoC and the flash (showing on 1st layer of Fig.~\ref{fig:MainXray}). They are respectively (light-green trace being the first) connected to the \texttt{I/O1}, \texttt{CLE} (command latch enable), \texttt{ALE} (address latch enable), \texttt{WE} (write enable), \texttt{RE} (read enable), \texttt{CE} (chip enable). Therefore, it is not possible to dump the memory from these test points. Indeed, only one IO line out of eight is exposed.

To extract all the data contained on the memory chip, it is thus necessary to desolder the chip with an infrared reworking station to access its pins. The process is using infrared light to heat the target component at the optimal reflow temperature~\cite{HECKMANN2018,BREEUWSMA2007}.
To facilitate the process, a backheater heats the whole board, using a thermal resistance. The backheater's purpose is to reduce the thermal difference between the chip to desolder and the board. This technique reduces the risk of fracturing the chip or the board. With this kind of process, it is possible to chip off the memory, dump it, then chip it on after reballing~\cite{HECKMANN2018}.

\begin{figure}[!b]
	\centering
    \includegraphics[width=0.95\textwidth]{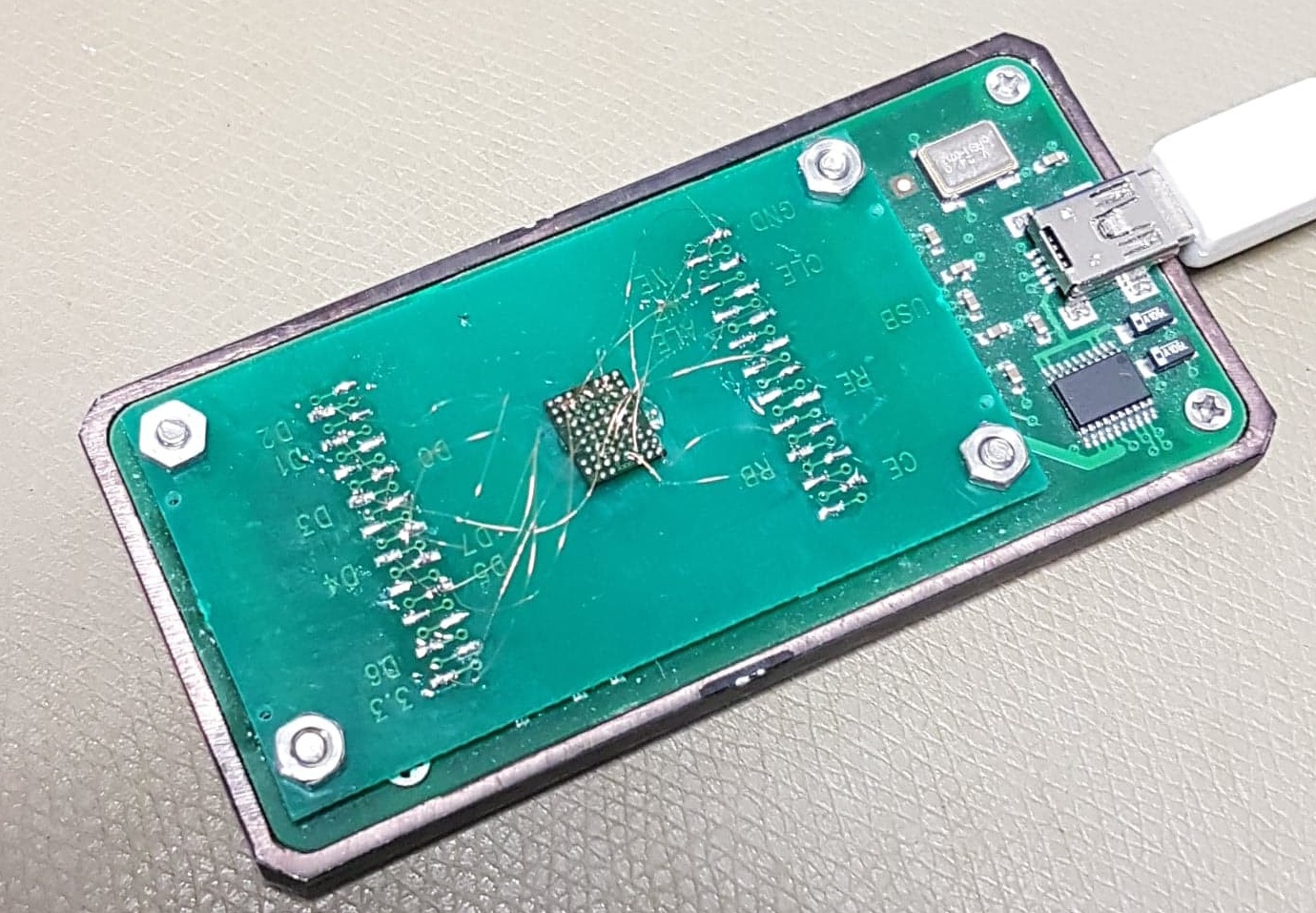}
    \caption{Optical view of the memory read-out assembly, with the wire-to-wire connection between the memory chip and a read-out board. The memory chip has a P-VFBGA67-0608-0.80-001 packaging, of which 17 pins corresponding the an ONFI interface.}
    \label{fig:desoldering}
\end{figure}

After the memory is chipped off, its contents are dumped. As there is no available socket for the pinout of the memory on the reader used, the chip is cabled with a wire-to-wire method (Fig.~\ref{fig:desoldering}).\footnote{Ideally, reballing the chip on a custom PCB is handier. However, it is often prohibitively expensive to order a unique PCB from third-party providers.} The chip is then fixed to an adapter board, while seventeen small coated wires (seven for controls, eight for Input/Output, and two for \texttt{VCC} and \texttt{GND}~\cite{ONFI}) are positioned between the signals of interest and the adapter. The dumping speed is reduced to minimize any error occurring during the process, taking 4 hours in total.

\section{Dumping the flash}
\label{sec:dumping}

The dumped \emph{raw image} is $285,212,672$ bytes (272MiB) long, segmented into 2176-byte \emph{memory pages}. Each memory page has 2048 \emph{data bytes} and 128 \emph{spare bytes} (similar devices usually feature 256 spare bytes). The spare bytes contain metadata like indication whether the memory page is damaged and shall not be used or error correcting codes, adding redundancy to the stored data in case of hardware errors.

According to the datasheet, the SoC can directly perform error correction. Further reverse engineering confirms the presence of hardware registers controlling the SoC's ECC capabilities. Comparison with other leaked Marvell datasheets points to the presence of a (17360, 16640, 97) Bose-Chaudhuri-Hocquenghem code~\cite{BOSE1960,HOCQUENGHEM1959} with 48 bits of error correction capability. However, the attempts at interpolating the polynomial using the Berlekamp-Massey algorithm~\cite{BERLEKAMP1968} were unsuccessful. In all likelihood, some non-linear operations are performed on the data before the computation of the syndrome. Those operations are detailed in the Marvell Armada 1500 Mini Plus' datasheet, which is not available online. Thence, another method is required to correct hardware errors based solely on the \emph{main image} obtained by removing the spare bytes from the raw image.

\begin{figure}[!hb]
    \includegraphics[width=\textwidth]{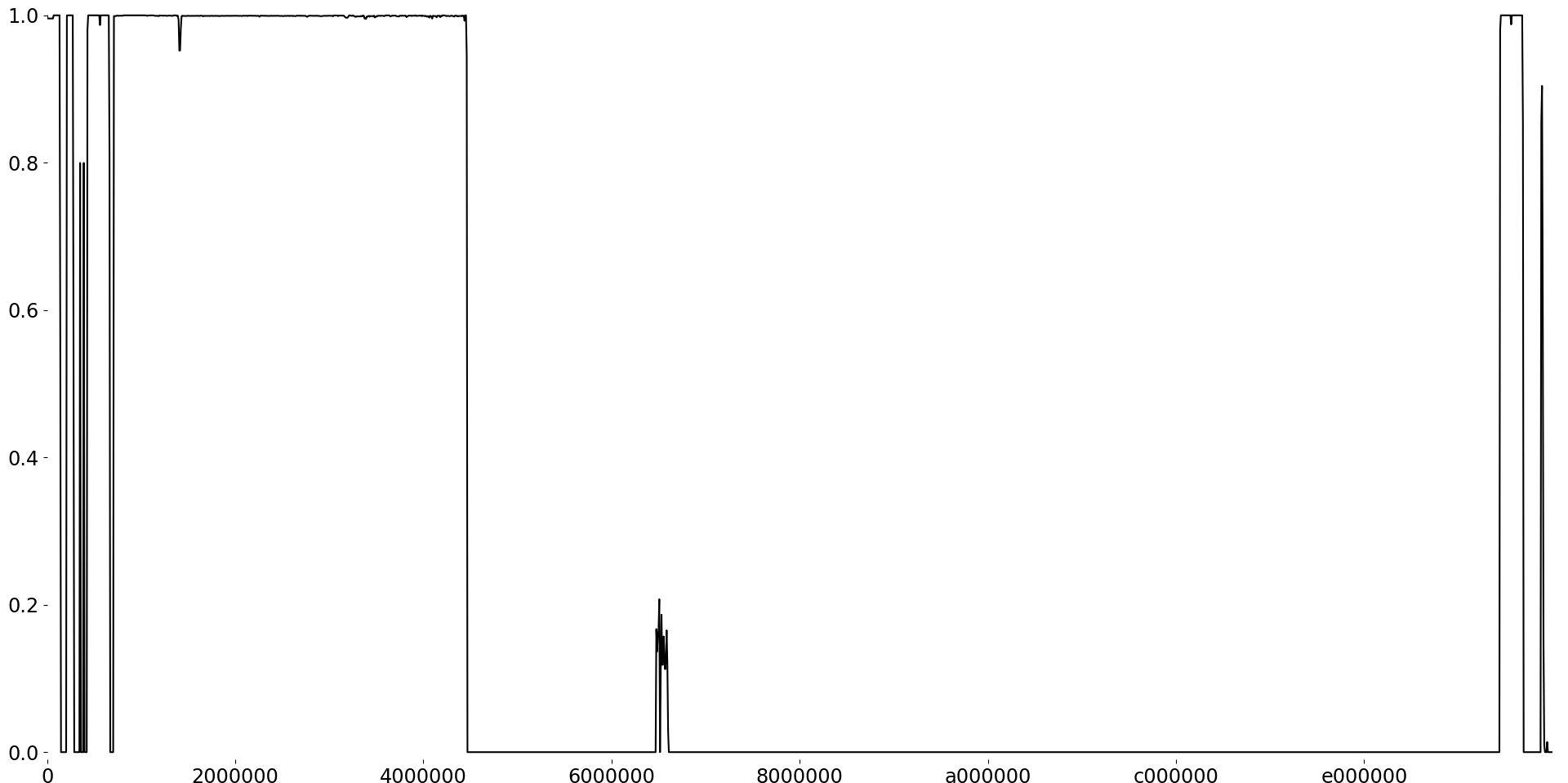}
    \caption{Shannon entropy (normalized) analysis of the main image using \texttt{binwalk}. The x-axis designates the offset relative to the start of the dump while the y-axis shows the entropy of a small region centered around a given offset.  }
    \label{fig:entropyanalysis}
    \centering
\end{figure}
\begin{figure}[!hp]
    \centering
\begin{tabular}{|l|l|r|}
  \hline
  start address & end address & \ctr{description} \\
  \hline
  \texttt{0x0      } & \texttt{0x120000 } & unknown \\
  \texttt{0x120000 } & \texttt{0x260000 } & unknown encrypted \\
  \texttt{0x260000 } & \texttt{0x360000 } & null bytes \\
  \texttt{0x360000 } & \texttt{0x4A0000 } & unknown encrypted \\
  \texttt{0x4A0000 } & \texttt{0x5A0000 } & null bytes \\
  \texttt{0x5A0000 } & \texttt{0x5C0000 } & SecureMonitor and bootloader \\
  \texttt{0x5C0000 } & \texttt{0x640000 } & null bytes \\
  \texttt{0x640000 } & \texttt{0x660000 } & SecureMonitor and bootloader \\
  \texttt{0x660000 } & \texttt{0x6E0000 } & null bytes \\
  \texttt{0x6E0000 } & \texttt{0xAB6840 } & [corrupt] \hfill mkbootimg zImage  \\
  \texttt{0xAB6840 } & \texttt{0xB60000 } & null bytes \\
  \texttt{0xB60000 } & \texttt{0x4780000} & [corrupt] \hfill SquashFS  \\
  \texttt{0x4780000} & \texttt{0x67A0000} & null bytes \\
  \texttt{0x67A0000} & \texttt{0x69A1A00} & unknown \\
  \texttt{0x69A1A00} & \texttt{0xF720000} & null bytes \\
  \texttt{0xF720000} & \texttt{0xFB18500} & [corrupt] \hfill android boot zImage \\
  \texttt{0xFB20000} & \texttt{0xFE20000} & null bytes- \\
  \texttt{0xFE20000} & \texttt{0xFEA1A00} & YAFFS overlayfs \\
  \texttt{0xFF20000} & \texttt{0xFF300C0} & crash dumps\\
  \texttt{0xFF300C0} & \texttt{0xFF34000} & ECC\\
  \texttt{0xFF34000} & \texttt{0xFFC0000} & null bytes \\
  \texttt{0xFFC0000} & \texttt{0xFFE0800} & Bad Block Table\\
  \texttt{0xFFE0800} & \texttt{0xFFFFFFF} & null bytes \\
  \hline
\end{tabular}
    \caption{General layout of the main image. }
    \label{fig:memlayout}
\end{figure}

A Shannon entropy analysis~\cite{SHANNON1948} of the main image in Fig.~\ref{fig:entropyanalysis}, highlights two types of segments, either of zero entropy (corresponding to null bytes), or of high entropy. Zooming onto the high entropy segments shows segments with entropy varying between 0.99980 and 0.99983 and segments with entropy varying between 0.998 and 1.000 with drops that go as low as 0.95. The first is characteristic of encrypted data while the latter has variations of entropy that characterize the absence of encryption, instead showing that the segments are compressed~\cite{HEFFNER2013,HEFFNER2013b}.

Running binwalk in signature analysis mode provides additional insight used to extract the layout detailed in Fig.~\ref{fig:memlayout}. Although all sections are not reversed, it is still possible to locate the critical elements used in the boot process of an Android device.

The boot process starts by executing the boot code contained in a
boot ROM located in the SoC. The boot code initializes the hardware and checks the integrity and authenticity of the bootloader located in the flash, before executing it. The bootloader checks the integrity and authenticity of the flash, and calls a small utility that decompresses the Linux kernel (known as the \emph{zImage}) into the RAM. The kernel then calls the \texttt{init} process (with PID 1). Additional steps initializing security elements (Secure Monitor, TrustZone, ...) are not detailed here.

The comparison of two encrypted sections of the flash shows minor differences that are incompatible with the avalanche effect of encryption. These differences can thus only occur after encryption, \emph{i.e.}, during storage. These memory errors are called \emph{bitflips}. Bitflips happen naturally in storage, although some methods aim at deliberately causing them. These bitflips allow to finely estimate the rate of bitflips coming only from data storage. Overall, there are 11 differences between both $1,310,720$ byte-long area, giving a proportion of one bitflip every $(2\times1310720)/11= 238312$ bytes. An informal preliminary analysis does not show any obvious pattern in the occurrence of bitflips. In the rest of this paper, a Bernoulli error model of parameter $p$ is used to more precisely quantify the bitflips. More formally, we assume that each bit of the flash flips following a sequence of \emph{independent, identically distributed} (IID) random variables whose probability distribution is a Bernoulli distribution of parameter $p$.

Although the measured bitflip rate seems reasonably low, it was enough to corrupt three important sections of the dump, namely the \texttt{mkbootimg} zImage, the SquashFS filesystem, and the android boot zImage. Indeed the first and the last are compressed with LZMA~\cite{PAVLOV2007}, while the second uses gzip~\cite{RFC1950,RFC1951,RFC1952}. Without correcting the bitflips, recovering the data is impossible. The two compression algorithms belong to two different families: LZMA is a stream compression algorithm, for which a bitflip corrupts the whole subsequent stream, while gzip uses a block compression algorithm, for which a bitflip impacts only the block in which it occurs.

As a general idea, given the sparsity of bitflips, it becomes realistically achievable for block compression algorithms to bruteforce one or two bitflips in each block until successful repair.
For stream algorithms, the stream is decompressed until an unrecoverable error is reached, from which a bitflip backward is bruteforced, greedily maximizing the length of successfully decompressed stream. However, the latter requires significant manual intervention to get the algorithm out of local minima, hampering its reproducibility to other use cases.
The rest of this paper only focuses on block compression, by detailing the process to repair the corrupted SquashFS filesystem, leaving stream compression for future work.

\section{Data recovery from the corrupt SquashFS dump}
\label{sec:repair}

\begin{figure}[!b]
	\centering
	\begin{tikzpicture}[scale=1.4]
		\draw(0,0.5) rectangle (8,-3);
		\draw(1.8,0.5) to (1.8,-1);
		\draw(4,0.5) to (4,-3);
		\draw(6,0.5) to (6,-1);
		\draw(3,0) to (3,-1);
		\draw(5,0) to (5,-1);
		\draw(0,0) to (8,0);
		\draw(0,-1) to (8,-1);
		\draw(0,-1.5) to (8,-1.5);
		\draw(0,-2) to (8,-2);
		\draw(0,-2.5) to (8,-2.5);

		\path (0,0.7) -- node (success) {128-bits} (8,0.7);
		\draw[<-,thick] (0,0.7) -- (success);
		\draw[->,thick] (success) -- (8,0.7);
		
		\node[text centered] at (0.9,0.25) {\texttt{0x73717368}};
		\node[text centered] at (2.9,0.25) {inode count};
		\node[text centered] at (5,0.25) {timestamp};
		\node[text centered] at (7,0.25) {block size};
		\node[text centered,text width=1.8cm] at (0.9,-0.5) {fragment entry count};
		\node[text centered,text width=1.2cm] at (2.4,-0.5) {compr-ession algo.};
		\node[text centered,text width=1cm] at (3.5,-0.5) {block size (log)};
		\node[text centered,text width=1cm] at (4.5,-0.5) {flags};
		\node[text centered,text width=1cm] at (5.5,-0.5) {id count};
		\node[text centered] at (7,-0.5) {version};
		\node[text centered] at (2,-1.25) {root directory inode};
		\node[text centered] at (6,-1.25) {archive size (bytes)};
		\node[text centered] at (2,-1.75) {id table start offset};
		\node[text centered] at (6,-1.75) {xattr id table start offset};
		\node[text centered] at (2,-2.25) {inode table start offset};
		\node[text centered] at (6,-2.25) {directory table start offset};
		\node[text centered] at (2,-2.75) {fragment table start offset};
		\node[text centered] at (6,-2.75) {export table start offset};
	\end{tikzpicture}
	\caption{Layout of a SquashFS superblock.}
	\label{fig:squashfstable}
\end{figure}
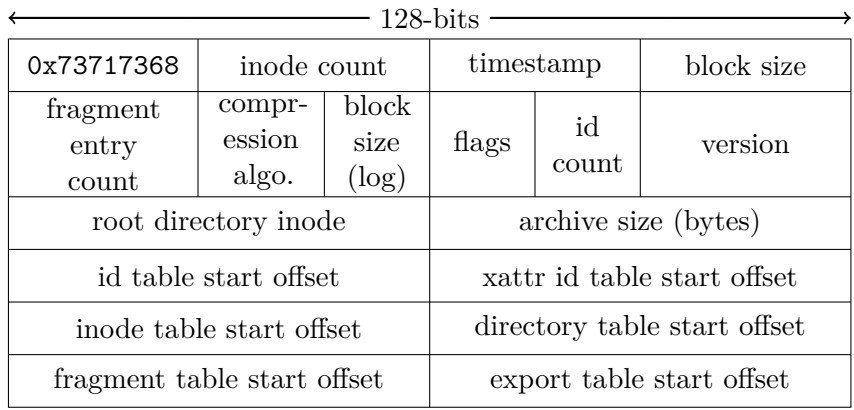

This sections details the steps taken to correct the bitflips on the corrupted SquashFS dump. SquashFS is a read-only filesystem, often used to store the operating system of embedded devices. Common configuration combines it with an other writable partition on same mounting point -- called the overlay filesystem -- whose files take precedence over the underlying SquashFS. This eases factory resets of embedded devices, as erasing the overlay filesystem reverts the device back to its original state, without any user-data stored therein.

A SquashFS image is divided in at most nine parts~\cite{SQUASHFS}, starting with a \emph{superblock}, whose layout is shown in Fig.~\ref{fig:squashfstable}. In the Google Home, the block size is defined to 128KiB, while the compression method is set to gzip -- the default and most common one. Each file or directory is referred to using an \emph{inode}, a special structure holding the file's metadata.

To improve storage efficiency, SquashFS compresses its inodes by packing them into \emph{metadata blocks} of size 8KiB, which are then compressed with gzip and stored in the \emph{inode table}. Similarly, each file is also compressed by splitting it into \emph{fragments} of at most 128KiB, which are then compressed using gzip.

The \texttt{squashfs-tools} utility fails at decompressing 204 out of the 920 fragments (22\%) of the filesystem. This amounts to 111 out of the 1139 files listed in the inode table. Fortunately, the inode table is not corrupted.

This information allows to refine the estimation of the bitflip rate $p$, by modeling the corruption of each fragment $i$ with a random variable $Y_i$ equal to zero if the fragment is not corrupted and one if the fragment is corrupted. Thus, $Y_i$ is a Bernoulli variable of parameter $1-(1-p)^{\textnormal{length}_i}$ (where $\textnormal{length}_i$ is the length of the fragment $i$).

The number of expected corrupted fragments can be estimated by summing all random variables $Y_i$. The expectation value is then equaled to the observed count of corrupted fragments (\emph{i.e.}, 204). This yields a $p$ equal to $5.03\times10^{-7}$ or equivalently, one bitflip every 248253 bytes.
\begin{align}
204 =& \: \mathbb{E}\left(\sum_{i=1}^{920} Y_i\right) = \sum_{i=1}^{920} \mathbb{E}(Y_i) = \sum_{i=1}^{920} 1-(1-p)^{\textnormal{length}_i} \label{eq:1} \\
p =& \: 5.03\times10^{-7}
\end{align}

Using Hoeffding's inequality~\cite{HOEFFDING1963}, the deviation from the expectation can be bounded as follows:
\begin{align*}
P\left(\left|\sum_{i=1}^{920} Y_i - \mathbb{E}\left(\sum_{i=1}^{920} Y_i\right)\right| \geq t\right) \leq 2\, \textnormal{exp} \left(-\frac{2\,t^2}{\sum_{i=1}^{920} 1 - 0}\right)
\end{align*}
Solving $t$ for a probability of $10^{-2}$ yields $t=50$. By replacing the right-hand side of Equation~(\ref{eq:1}) by $204\pm t$, this gives that the bitflip rate $p$ is in the range $[\,3.66 \times 10^{-7} ; 6.53 \times 10^{-7}\,]$  with probability $99\%$ or equivalently, one bitflip every $[\,191255 ; 341937\,]$ bytes. This range is consistent with the preliminary manual analysis, confirming the hypothesis of an identical bitflip rate across all sections of the flash.\footnote{Using the Bienaymé-Chebyshev inequality (with variance equal to $160$) yields less precise bounds for same probability $10^{-2}$. }

The gzip compression method~\cite{RFC1952}, based on the DEFLATE algorithm, is a wrapper around zlib compressed data~\cite{RFC1950}, itself concatenating the DEFLATE compressed~\cite{RFC1951} data to some additional metadata. As a crude approximation, each \emph{compressed fragment} can be summarized as in Fig.~\ref{fig:zliblayout}:

\begin{figure}[ht!]
\centering
  \begin{tikzpicture}[scale=1]
\draw(-0.5,0) rectangle (10,-1);
\draw(1,0) to (1,-1);
\draw(3,0) to (3,-1);
\draw(4,0) to (4,-1);
\draw(6,0) to (6,-1);
\draw(8,0) to (8,-1);

\node[text centered] at (0.25,-0.5) {header};
\node[text centered,text width=1.5cm] at (2,-0.5) {deflate block};
\node[text centered] at (3.5,-0.5) {...};
\node[text centered,text width=1.5cm] at (5,-0.5) {deflate block};
\node[text centered,text width=1.5cm] at (7,-0.5) {deflate block};
\node[text centered,text width=1.5cm] at (9,-0.5) {checksum};
  \end{tikzpicture}
    \caption{Layout of a zlib compressed fragment.}
    \label{fig:zliblayout}
\end{figure}

The checksum is computed on the decompressed data using the Adler-32 algorithm. Beyond the fact that Adler-32 is not meant for error correction, one bitflip on compressed data may result in numerous bitflips in the decompressed data, making any attempt for correction after decompression impractical.

Instead, a blackbox approach looks more adequate, by modeling the Adler-32 check as an oracle, querying it with a \emph{repair candidate}, a compressed fragment to which a bitflip is applied, and determining if the repair candidate is \emph{valid} or \emph{invalid}. A valid candidate can then be decompressed using \texttt{gzip} to produce a \emph{target candidate}.

\subsection{Generating target candidates}
\label{sec:targetgen}
To refine this strategy, a finer estimation of the number of bitflips in the corrupted fragments must be performed. Using the previously computed rate $p$, it is now possible to model the number of bitflips in each fragment $i$ using a Binomial variable $Z_i$ of parameters ($\textnormal{length}_i, p$). It is then possible to compute the expected number of fragments that have $k$ bitflips as follows, where $\delta$ is the Kronecker delta, \emph{i.e.}, $\delta_{Z_i}^k = \begin{cases}1, Z_i = k \\ 0, Z_i \neq k\end{cases}$
\begin{align}
\mathbb{E}\left(\sum_{i=1}^{920} \delta^{k}_{Z_i}\right) = \sum_{i=1}^{920} \mathbb{E}(\delta^{k}_{Z_i})  =  \sum_{i=1}^{920} \binom{\textnormal{length}_i}{k} p^k \, (1-p)^{\textnormal{length}_i - k}
\end{align}

Out of the 204 corrupted fragments, $175.22$ are thus expected to have a single bitflip, $25.78$ to have 2 bitflips, and $2.75$ to have 3. Fragments with more than 4 bitflips are nearly inexistent. In what follows, reparations focus only on single and double bitflips.

To begin with, corrupted fragments are tentatively repaired using a single bitflip error model. For each of the 204 corrupted compressed fragments $f$, its associated set of repair candidates $C_f$ is generated by mutating $f$ with a single bitflip. Three criteria are used to discriminate repair candidates and generate the target candidates set $T_f$:
\begin{enumerate}[nolistsep,before=\vspace{0.5\baselineskip},after=\vspace{0.5\baselineskip}]
	\item One of the deflate blocks of the fragment is corrupted and cannot be inflated.
	\item The Adler-32 of the decompressed data is not correct.
	\item The length of the decompressed fragment exceeds 128KiB.
\end{enumerate}

The repair is said to be successful if the cardinality of $T_f$ is one (\emph{e.g.}, there is exactly one target candidate). The repair candidate generation can be optimized. If the number of bytes $n$ read from the compressed fragment $f$ when trying to decompressing it using zlib is less than the length of $f$, this means than a bitflip occurs in the first $n$ bytes of $f$, allowing to reduce the set of repair candidates to only the ones that have a bitflip in their first $n$ bytes. This optimization significantly cuts the search interval for 5 out of 204 fragments, while for all others, the full input is read.

\subsection{Additional inode table based oracle}
\label{sec:additionalfilter}

For fragments having multiple repair candidates, it is possible to reduce the number of valid target repair candidates by using the decompressed fragment length. As the length of each file is stored in the inode table, only target candidates whose lengths after decompression are compatible with their associated file lengths are kept. For simplicity purposes, a trial and error approach is performed, by assuming that all remaining fragments have length 128KiB.

After decompressing the SquashFS, the length of each file is checked against its length in the inode table. In our case, the lengths match for all files, validating the initial assumption about the length of the 29 fragments. Had one file been of incorrect length, only combinations of target candidates whose sum of lengths are equal to the length of the file should have been kept. This can be implemented as a variant to the subset sum problem~\cite{MARTELLO1990}. This additional filter allows to reduce the number of target candidates significantly, as shown in the last column of Appendix~\ref{app:listing}.

\section{Results and Discussion}
\label{sec:results}

The 1-bitflip repair process takes 73 minutes on an Intel i7-8700 machine (6c/12t). Out of the 204 corrupted fragments, 172 of them have a target candidates set of cardinality one. Out of the remaining 32, 29 can't be repaired with a single bitflip (cardinality of target candidates is zero), while 3 have multiple targets. Overall 102 out of the 111 corrupt files of the filesystem have been repaired using a 1-bitflip error model. The remaining nine files are listed in Appendix~\ref{app:listing}, along the name of the fragments associated to the number of target candidates for each fragment.

For the remaining 29 fragments, a double-bitflip error model (all combinations of 2 bitflips in a fragment) is used to generate the repair candidates. The process takes several months per fragment on an Intel i7-8700 machine. Execution time is proportional to the cube of compressed fragment's size. All fragments have been repaired: the smallest takes 22 days, while the biggest takes more than a year. The task is distributed across 40 computers to reduce execution time to a week per fragment. The number of target candidates for each fragment ranges from 1 up to 40717 (Appendix~\ref{app:listing}). The next paragraphs details how to handle multiple target candidates.

\subsection{Merging multiple target candidates}

\begin{figure}[!b]
	\centering
	\includegraphics[width=\textwidth]{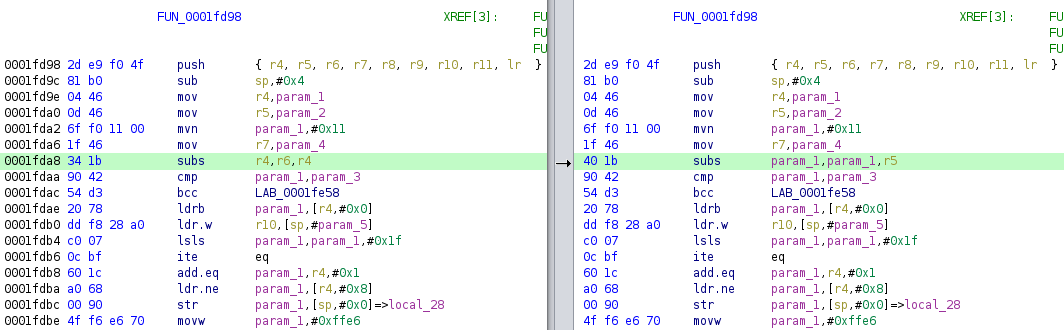}
	\caption{Differences arising in the file \texttt{/bin/bluetoothtbd} from the two target candidates for fragment \texttt{heNYlKQRQ8pfXf3Z3PPPrTiGCnkbaLLai2lenD8qRzA} . The left candidate presents an instruction (in green) using an undefined register \texttt{r6}, while the right candidate seems valid. Thence, the left candidate is not valid, and is discarded.}
	\label{fig:ghidra}
\end{figure}

Two merging methods are used to produce the final result. If the number of target candidates for a fragment is small (typically 2), one file per target candidate is generated, and manually analyzed using \texttt{ghidra}'s~\cite{GHIDRA} binary diffing tool. The tool highlights any difference between the target candidates, to eliminate the aberrant ones. This process is entirely manual, time consuming, and needs to be performed from scratch for each fragment. As an example, Fig.~\ref{fig:ghidra} details this process for the fragment \texttt{heNYlKQRQ8pfXf3Z3PPPrTiGCnkbaLLai2lenD8qRzA} (complete listing is available in Appendix~\ref{app:listing}), by comparing two versions of the binary executable file \texttt{/bin/bluetoothtbd}. There are two target candidates, each corresponding to a binary file. The tool then pinpoints the differences between the two files, and shows the binary code associated. The candidates that feature code with aberrant behaviors are then discarded. The results presented in Section~\ref{sec:recovery} do not take into account fragments repaired using this method, as its efficiency heavily depends on human expertise to discriminate the targets, which cannot be accurately measured.

The other method merges indiscriminately all target candidates by using three-valued logic, where a bit is of value true (respectively false) if and only if this bit is equal to true (respectively false) in all target candidates, otherwise it is of value indeterminate. For this, the \texttt{sasquatch} utility is patched to generate two variants of each file, one in which all indeterminate bits are set to true, and another in which all indeterminate bits are to false. Comparing the two output files using a binary diffing tool like \texttt{ghidra} or \texttt{bindiff} as previously allows to highlight any section that has not been soundly repaired.

\subsection{Recovery ratios}
\label{sec:recovery}

\begin{figure}[!h]
\centering

\begin{tabular}{|l|l|l|}
	\hline
	Repair method & Bits (ratio) & Bytes (ratio) \\\hline
	\texttt{squashfs-tools} & $78,450,840$ ($8.32\%$) & $9,806,355$ ($8.32\%$) \\\hline
	Bitflip repair & $939,968,046$ ($99.68\%$) & $117,068,734$ ($99.31\%$) \\\hline
	Total &  $942,990,616$ & $117,873,827$ \\\hline
\end{tabular}
\caption{Recovery ratios at bit-level, byte-level and file-level granularity before and after performing bitflip repair.}
\label{table:results}
\end{figure}

As a reminder from previous sections, currently available \texttt{squashfs-tools} utility manages to decompress only $1028$ files out of the $1139$ files of the inode table, amounting to only $9,806,355$ out of a total of $117,873,827$ bytes of decompressed data, which represents a recovery ratio of $8.32\%$. After the repair process, there are 3022570 indeterminate bits. This amounts to $0.32\%$ of corrupted data at a bit-level granularity. If considering bytes in which at least one indeterminate bit as corrupted, this amounts to 805093 corrupted bytes or a ratio of $0.68\%$.

Overall, the repair process manages to increase the ratio of recovered data from $8.32\%$ to $99.68\%$, as shown in Fig.~\ref{table:results}, significantly helping any investigative effort, even though this is still significantly below the repair capabilities of the Marvell Armada SoC. Ideally, the ECC algorithm should always be made public by the manufacturer. 
Indeed, investigators would benefit greatly from having access to a description of the ECC algorithms employed by these devices with no impact on the overall security. Such algorithms are usually not proprietary and revealing them does not introduce security vulnerabilities. 

\section{Conclusion}
\label{sec:conclusion}
This paper describes the steps taken at identifying the various components of a Google Home in order to extract its firmware. The efforts at recovering the main SquashFS filesystem were hindered by the use of undisclosed non-linear operations before the computation of the BCH error correcting code. To alleviate this, this paper presents, to the best of the authors' knowledge, an original method aiming at repairing gzip compressed data, by leveraging residual redundancy embedded in the data. Under the hypothesis of a low-bitflip rate, most of the data can be recovered, at the cost of several months of computation.

This oracle based approach may be easily reproduced to other corrupted SquashFS filesystems. As a future work, this methodology can be extended to other compression algorithms, specifically stream compression like \texttt{lzma}. Preliminary research efforts aiming at the recovery of the \texttt{zImage} yield promising results that must generalized be to be reproducible with minimal human intervention.

\section*{Declaration of competing interests}
The authors declare that they have no known competing
financial interests or personal relationships that could have
appeared to influence the work reported in this paper.

This research did not receive any specific grant from funding agencies in the public, commercial, or not-for-profit sectors. This work was performed when R\'emi G\'eraud was at \'Ecole normale supérieure.

\clearpage
\printbibliography[]

\clearpage
\appendix

\section{Source code and artifact}
The source code of the SquashFS repair utility, as well as the artifact to reproduce the statistics provided in this paper, are available on the following
url: \url{https://github.com/enssec/squashfs_bitflip_repair/}.

\section{Fragments with multiple target candidates}
\label{app:listing}

The following table lists all fragments (identified by their hash) with multiple target candidates and lists respectively the number of target candidates (Sec.~\ref{sec:additionalfilter}) of size 128 KiB, how many indeterminate bits, and how many bytes in the fragment have at least one indeterminate bit.

\begin{center}
\begin{adjustbox}{angle=270}
\footnotesize
\noindent\begin{tabular}{|l|c|c|c|c|}
	\hline
	File name & fragment (hash) & $\#$targets & \multicolumn{2}{c|}{$\#$indet. bits / bytes} \\
	\hline
	\verb!/bin/bluetoothtbd! & \verb!heNYlKQRQ8pfXf3Z3PPPrTiGCnkbaLLai2lenD8qRzA! 
	& 2 & 10 & 3 \\
	\hline
	\verb!/bin/wpa_supplicant! & \verb!92bEmBqIKN9dGW%AtlCg3fWYGgsYhTJ5DHuenvz8dbY! 
	& 105 & 3961 & 1136 \\
	\hline
	\verb!/boot/recovery.img! & \verb!256HgEHVU@6U0uNouwruyGDWO%2jTniABl%NEEhWKRI! 
	& 40717 & 80984 & 59421 \\
	& \verb!1QCNtpab0aCWKV68Ydo2IWnoo5IqLN4zYy3vezSCdzE! 
	& 2 & 2 & 2\\
	\hline
	\verb!/chrome/assets/cast_shell.pak! & \verb!Sh43xfhLF3@Remh2coYSxiChVxt2SqW0iyLw%o9ApGs! 
	& 2 & 247123 & 58624 \\
	\hline
	\verb!/chrome/cast_shell! & \verb!wC8p%RucvcX4EDStclRmSsO5a7CVA1W6WHrUFcrJKVo! 
	& 1354 & 30713 & 8456
	\\ & \verb!JOw2De0T4V3G@0vSQFrp6Ie6%7myxs%iHJjQpMl0lTI! 
	& 3 & 999 & 245
	\\ & \verb!D2dCzirIrlRGNTIKjCMn2sIJyquiLNs8guKlHCQNfRU! 
	& 394 & 13671 & 4033
	\\ & \verb!wOhHT5XDso@EZ7kkp39lcyEOyNKXo1BC7DsT2h6EUZc! 
	& 3 & 400268 & 104706
	\\ & \verb!fstWXSWTHt2jVMNr0C1lU@qNAEntfS@BQD%XV3WjDCE! 
	& 124 & 10501 & 2778
	\\ & \verb!GVioeSeF8NRKMMULiTs0Ns8xQr2J9ytABwuIRSwbquI! 
	& 2 & 296963 &73781
	\\ & \verb!4jQJy1mkwBhbDEuGGH28WdzPSWswwaTFkq5fhiyujVE! 
	& 414 & 19355 & 5076
	\\ & \verb!b1VT5jiuhOqxRX%xa7INu52iYt%t9bpL2aVe7nfp7os! 
	& 59 & 8163 & 2224
	\\ & \verb!r1zrx7VtnidJz6ohWtwyDILTc2x1h4cZ3Y%yVcg8@Zg! 
	& 8 & 387421 & 99010
	\\ & \verb!JHMZLlzG%FqGyCYFV2fVxL3@lqLZq1e3lKsU7%0ZtxA! 
	& 2 &19 &7
\\
	\hline
	\verb!/chrome/icudtl.dat! & \verb!Wnu5X5DpQtyOHAgAdOJOnZ0@k8xUVu8w5Yc7TLatcY4! 
	& 2337 & 257725& 64282
	\\ & \verb!i3LU8g00MdGpitYFOsNAVaI4M%Hjb2lJ6Odhjwnxgs4! 
	& 6 & 430705 & 94562
	\\ & \verb!r2DiXJjXryJsggIHr1Ca4HXGdgXg3j3Tnloza9bynCM! 
	& 2 & 6144 & 1558
\\
	\hline
	\verb!/chrome/lib/libassistant.so! & \verb!t2hs5Z71xQbRDW1LjNMPHvtyLoP%jC3voPG%R@nhD1c! 
	& 2 & 18050 & 4935
	\\ & \verb!Km21LAN6wnDRqmpBA@RMXTUQmaQ9fSHhXDXWrR9cHoM! 
	& 24709 & 93454 & 39408 \\
	\hline
	\verb!/chrome/libffmpeg.so! & \verb!WHi3paAUS9t6DdQ2Ka2CxkcRyTLhnTiSztpftJ%3kzs! 
	& 5 & 318888 & 81323
	\\
	\hline
	\verb!/lib/libfreeblpriv3.so! & \verb!nD4ME7GTBAsBjKfVIC7YprIeYNT%4Kru1%4RDinT3aA! 
	& 6 & 397451 & 99523
	\\
	\hline
\end{tabular}
\end{adjustbox}
\end{center}

\end{document}